\documentclass[prd,reprint,onecolumn,superscriptaddress,tightenlines,nofootinbib,eqsecnum]{revtex4-2}

\usepackage{amsmath}
\usepackage{cancel}
\usepackage{amsfonts}
\usepackage{amssymb}
\usepackage{bm}
\usepackage[colorlinks]{hyperref}
\usepackage{mathrsfs}
\usepackage{graphicx}
\usepackage{multirow}
\usepackage{empheq}
\usepackage{ulem}
\usepackage{tensor}
\usepackage{tabularx}
\usepackage[usenames]{color}
\definecolor{darkgreen}{rgb}{0,0.6,0}
\hypersetup{urlcolor=darkgreen, citecolor=darkgreen}
\usepackage{cleveref}
\usepackage{stmaryrd}
\newcolumntype{Y}{>{\centering\arraybackslash}X}

\allowdisplaybreaks

\DeclareSymbolFontAlphabet{\mathrsfs}{rsfs}
\DeclareMathAlphabet{\mathcal}{OMS}{cmsy}{m}{n}

\usepackage{parskip}
\definecolor{darkgreen}{rgb}{0,0.5,0}

\hypersetup{
    unicode=false,          
    pdftoolbar=true,        
    pdfmenubar=true,        
    pdffitwindow=false,     
    pdfstartview={FitH},    
    pdftitle={My title},    
    pdfauthor={Author},     
    pdfsubject={Subject},   
    pdfcreator={Creator},   
    pdfproducer={Producer}, 
    pdfkeywords={keyword1} {key2} {key3}, 
    pdfnewwindow=true,      
    colorlinks=true,       
    linkcolor=red,          
    citecolor=cyan,        
    filecolor=magenta,      
    urlcolor=darkgreen,           
    linktocpage=true}

\newcommand{\be}{\begin{equation}}
\newcommand{\ee}{\end{equation}}

\newcommand{\mutp}{\widetilde{\mu}^{(2)}_{+}}
\newcommand{\mutm}{\widetilde{\mu}^{(2)}_{-}}
\newcommand{\sigmatp}{\widetilde{\sigma}^{(2)}_{+}}
\newcommand{\sigmatm}{\widetilde{\sigma}^{(2)}_{-}}
\newcommand{\muthp}{\widetilde{\mu}^{(3)}_{+}}
\newcommand{\di}{\mathrm{i}}
\newcommand{\e}{\mathrm{e}}
\newcommand\calO{{\mathcal{O}}}
\newcommand{\dd}{\mathrm{d}}

\newcommand{\nn}{\nonumber}
\newcommand{\fid}{\dot{\phi}}
\newcommand{\rd}{\dot{r}}

\newcommand{\tmass}{M}
\newcommand{\ADM}{\mathcal{M}}
\newcommand{\dI}{\mathrm{I}}
\newcommand{\dJ}{\mathrm{J}}
\newcommand{\dW}{\mathrm{W}}
\newcommand{\dU}{\mathrm{U}}
\newcommand{\dV}{\mathrm{V}}
\newcommand{\dM}{\mathrm{M}}
\newcommand{\dS}{\mathrm{S}}

\newcommand{\etidal}{\epsilon_\text{tidal}}
\newcommand{\et}{e_t}

\definecolor{mygreen}{rgb}{0,0.8,0}

\usepackage{ulem}
\allowdisplaybreaks

\usepackage{etoolbox}
\makeatletter
\patchcmd{\frontmatter@abstract@produce}
  {\vskip200\p@\@plus1fil
   \penalty-200\relax
   \vskip-200\p@\@plus-1fil}
\makeatother

\begin{document}

\title{Adiabatic tides in compact binaries on quasi-elliptic orbits: \\Radiation at the second-and-a-half relative post-Newtonian order}

\author{Quentin \textsc{Henry}}
\affiliation{Departament de Física, Universitat de les Illes Balears, IAC3 – IEEC, Crta. Valldemossa km 7.5, E-07122 Palma, Spain}

\date{\today}

\begin{abstract}
We compute the gravitational fluxes and waveform for eccentric compact binaries including matter effects through adiabatic tidal interactions within the post-Newtonian approximation. The computations are performed at the relative 2.5PN order. Using the dynamics derived in the companion paper, we first derive the radiated energy and angular momentum, from which we deduce the equations describing the secular evolution of the orbital elements. We numerically solve for the secular dynamics for various systems. We find that the eccentric corrections to tidal terms induce a dephasing that could potentially be detectable in some regions of the parameter space of gravitational wave sources. Finally, we compute the amplitude of the strain, decomposed in spin-weighted spherical harmonics. Besides the memory contributions that are left for future works, we provide the amplitude modes containing the instantaneous, tail and post-adiabatic corrections expanded to the twelfth order in eccentricity. All relevant results are provided in an ancillary file.
\end{abstract}


\maketitle

\section{Introduction}\label{sec:intro}

The recent release of the gravitational wave (GW) source catalog for the first part of the fourth observation run (known as O4a)~\cite{LIGOScientific:2025slb}, by the LIGO-Virgo-KAGRA collaboration (LVK), has brought the total number of detected GW events to over 200 and further solidified the field of GW astronomy. In detecting such events, the LVK has used several fast waveform generators. For parameter estimation in O4a these predominately included several phenomenological (Phenom) models~\cite{LIGOScientific:2025yae}, \texttt{IMRPHENOMNSBH}~\cite{Thompson:2020nei}, \texttt{IMRPHENOMPV2\_NRTIDALV2}~\cite{Dietrich:2019kaq}, \texttt{IMRPHENOMXO4A}~\cite{Hamilton:2021pkf, Thompson:2023ase} and \texttt{IMRPHENOMXPHM\_SPINTAYLOR}~\cite{Colleoni:2024knd}; effective-one-body models (EOB), \texttt{SEOBNRV4\_ROM\_NRTIDALV2\_NSBH}~\cite{Matas:2020wab} and \texttt{SEOBNRV5PHM}~\cite{Ramos-Buades:2023ehm}; as well as the numerical relativity surrogate (NRSurrogate), \texttt{NRSUR7DQ4}~\cite{Varma:2019csw}. Apart from the NRsurrogate, which interpolates known numerical relativity waveforms over a particular part of the parameter space, all other waveform models incorporate post-Newtonian (PN) expressions to inform the inspiral part of the waveform. Thus, the post-Newtonian framework not only finds itself still necessary for many GW detections, but due to its analytical nature, it also enables insights into the physics of the systems. 

Among the numerous detected compact binaries, a few of them involved at least one neutron star (NS)~\cite{LIGOScientific:2017vwq, LIGOScientific:2021qlt, LIGOScientific:2024elc}. The LVK now begins to power down for planned upgrades, which will make the detectors more sensitive to NS systems (see Table 1 of \cite{LVK:2025T2400403}). The expected increase in sensitivity for the upcoming fifth LVK observing run (O5) will allow for an order of magnitude increase in detections of both binary neutron stars (BNS)~\cite{Shah:2023ozh} and neutron star-black hole (NSBH)~\cite{Colombo:2023une} systems, with also the prospect for larger signal-to-noise ratios (SNRs) later in the signal. The possibility of seeing such an effect grows significantly when looking at third-generation ground-based detectors, like Einstein Telescope~\cite{ET:2025xjr}, which can have significant implications in high energy physics.

This project has notably been motivated by the event GW200105~\cite{LIGOScientific:2021qlt}, whose signal came from a NSBH system which entered the detector band with an eccentric motion. The eccentricity at the 20Hz frequency has been estimated to be be roughly $\sim 0.13$~\cite{Fei:2024ruj,Morras:2025xfu,Planas:2025plq,Kacanja:2025kpr,Jan:2025fps,Tiwari:2025fua}. However, these parameter estimation studies used waveform models  without including finite size effects. Today, the waveform models that describe such effects through tidal interactions are  \texttt{TEOBResumS-Dalí}~\cite{Gamba:2023mww,Albanesi:2025txj}, \texttt{SEOBNRv5THM}~\cite{Haberland:2025luz}, \texttt{NRTidalv3}~\cite{Abac:2023ujg}, \texttt{IMRPhenomXPHM\_NSBH}~\cite{IMRPhenomXPHMNSBH} or the so-called reduced-order models \cite{Purrer:2014fza,Lackey:2016krb,Lackey:2018zvw}. Among these, only \texttt{TEOBResumS-Dalí} is able to account for both eccentricity and tides simultaneously, although some PN information regarding eccentric tidal terms in the radiation reaction force is not included due to the lack of knowledge in the PN literature. Ref.~\cite{Huez:2025npe} discusses systematic biases arising when analyzing BNS signals with inadequate waveform models. The purpose of this paper is to fill this gap.

Within the PN approach, different physical effects, such as spins, finite size, black-hole absorption... are included through an effective matter action coming from effective field theory. One can later specify the motion and assume quasi-circular orbits or more general ones such as an eccentric or spin precessing in the case of misaligned spins. In the companion paper, called Paper~I~\cite{paperI}, we gave an overview of the PN literature treating with the conservative and radiative dynamics on eccentric motion regarding various physical effects, point-particle~\cite{DamourDeruelle1,
damour1985general,Damour:1988mr,Schafer:1993pkg,Memmesheimer:2004cv,Boetzel:2017zza,Cho:2021oai,Trestini:2025yyc}, spins~\cite{Damour:2004bz,Konigsdorffer:2006zt,Tessmer:2010hp,Klein:2010ti,Tessmer:2012xr,Samanta:2022yfe,Cho:2022syn,Henry:2023tka}, electromagnetic interaction~\cite{Henry:2023guc,Henry:2023len}. But none before Paper~I treated rigorously adiabatic tides on non quasi-circular orbits. 
In the radiative sector, the radiated fluxes for point-particles at the 3PN order, both in modified harmonic and ADM coordinates, have been derived in~\cite{Arun:2007sg,Arun:2009mc,Arun:2007rg}. It has been complemented to the same PN order with aligned spin contributions in~\cite{Henry:2023tka}. The secular evolution of the orbital elements are derived from these fluxes using balance equations for both non-spinning and spinning contributions. The amplitude of the GW strain has been tackled for non-spinning binaries at 3PN in the series of papers~\cite{Mishra:2015bqa,Boetzel:2019nfw,Ebersold:2019kdc}, the aligned spin contributions at 3PN in~\cite{Henry:2023tka} and precessing spins in~\cite{Blanchet:2006gy,Faye:2006gx,Arun:2008kb,Klein:2010ti,Buonanno:2012rv,Marsat:2013caa,Klein:2021jtd}. Note that the amplitude derived in~\cite{Boetzel:2019nfw,Ebersold:2019kdc,Henry:2023tka} was provided using an eccentricity expansion to the sixth order. Finally, finite size effects were also broadly studied within the PN or post-Minkowskian framework, through tidal interactions, see the non-exhaustive list~\cite{Flanagan:2007ix,Vines:2010ca,Damour:2012yf,Steinhoff:2016rfi,Banihashemi:2018xfb,Abdelsalhin:2018reg,HFB19,HFB20a,HFB20b,Cheung:2020sdj,Kalin:2020lmz,Mougiakakos:2022sic,Mandal:2023hqa,Patil2024,Bernard:2023eul,Dones:2024odv,Dones:2025zbs}, but not in the case of bound eccentric orbits. Most of these works used the adiabatic (or static) tides approximation, in which tidal effects are parametrized by Love numbers characterizing the deformability of a compact object with respect to an external tidal field.

The aim of the present paper is to derive the eccentric corrections to the tidal terms in the radiative sector. It is the part of the series of works~\cite{HFB19,HFB20a,HFB20b,Dones:2024odv,paperI}. To briefly summarize, we have considered the matter action~\eqref{eq:Smatter} describing adiabatic tides at next-to-next-to leading order (NNLO). We have derived the equations of motion and conserved quantities of the system~\cite{HFB19,HFB20b} for a general motion. Then, we derived the radiated energy flux and the phasing on quasi-circular orbits up to relative 2.5PN~\cite{HFB20a}. The computations to that PN order have later been extended to the GW amplitude on quasi-circular orbits in~\cite{Dones:2024odv}, which required more PN information. Finally, the present work computes the dynamics and GW radiation in the case of an eccentric binary. In Paper~I~\cite{paperI}, we focused on the conservative and radiative dynamics at relative 2.5PN, which has been derived employing a quasi-Keplerian parametrization based the conserved quantities derived in~\cite{HFB19} and the 2.5PN equations of motion of~\cite{Dones:2024odv}. In the present paper, we focus on the GW radiation, \textit{i.e.}, we derive the radiated energy and angular momentum, from which we deduce the secular evolution of the orbital elements and then compute the amplitude of the GW strain decomposed in spin-weighted spherical harmonics. The amplitude is expanded to the twelfth order in eccentricity.

The paper is organized as follows. In~\Cref{sec:formalism}, we recall the matter action that is considered, then we present brielfy the PN-MPM formalism and summarize the results of Paper~I concerning the motion including adiabatic tides. In~\Cref{sec:fluxes}, we derive the radiated energy and angular momentum. We first focus on the instantaneous part, then we compute the tail part of the fluxes using an eccentricity expansion, which we later resum. This allows to derive the total fluxes that are employed to derive the secular evolution equations of the orbital elements. In~\Cref{sec:modes}, we derive the amplitude modes of the waveform neglecting memory effects. 
In~\Cref{sec:num}, we numerically integrate the secular evolution of the orbital elements and discuss the potential importance of the eccentric corrections to the tidal terms on the phasing.
In Appendix~\ref{app:Ikn}, we derive general formulas for orbit average integrals  and in Appendix~\ref{app:lengthy} we display some lengthy results. All relevant results are provided in an ancillary file~\cite{SuppMaterial2} as a \textit{Mathematica} notebook. Its detailed content is given in the Conclusion Section~\ref{sec:ccl}.

\section{General formalism, recalls on previous works}\label{sec:formalism}

\subsection{Notations and conventions}\label{subsec:notations}

In this paper,\footnote{Greek tensor indices are four-dimensional $\mu=0,1,2,3$ and latin indices stand for spatial coordinates, \textit{i.e.} $i=1,2,3$ and the multi-index notation is $L=i_1\dots i_\ell$. The symmetric trace-free (STF) operator is noted by $\langle\dots\rangle$ around indices. The Levi-Civita tensor is noted $\epsilon_{ijk}$ with the convention $\epsilon_{123} = 1$.} we consider a binary system of compact objects within general relativity. Both objects, noted $A=1,2$, of mass $m_A$ are tidally interacting, where the matter action is written below. The constants $G$ and $c$ are respectively the Newtonian gravitational constant and the speed of light in vacuum. The total mass is denoted $M = m_1 + m_2$ with $m_1 \geq m_2$, $\nu=m_1 m_2/\tmass^2$ is the symmetric mass ratio, $\delta = (m_1-m_2)/\tmass = \sqrt{1-4\nu}$ is the normalized mass difference. Positions and velocities in the center-of-mass frame are $\bm{x} = \bm{y}_1-\bm{y}_2$, $\bm{v} = \dd \bm{x}/\dd t$ which allows to define the separation $r = |\bm{x}|$ and $\bm{n} = \bm{x}/r$. Thus the relative velocity is given by $\bm{v} = \rd \bm{n} + r\fid \bm{\lambda}$, where $\phi$ is the phase angle.

Tidal effects are modeled from effective field theory, in which we include the adiabatic tidal mass quadrupole and octupole interactions, as well as the current quadrupole, more details are given in~\cite{Damour:1990pi,Damour:1991yw,Bini:2012gu}. The matter action that we use reads
\begin{equation}\label{eq:Smatter}
S_\text{matter} = \sum_{A} \int \dd \tau_{A} \left[ -m_A c^2 + \frac{\mu_{A}^{(2)}}{4} G^{A}_{\mu\nu}G_{A}^{\mu\nu} + \frac{\sigma_{A}^{(2)}}{6c^{2}}H^{A}_{\mu\nu} H_{A}^{\mu\nu} + \frac{\mu_{A}^{(3)}}{12} G^{A}_{\lambda\mu\nu} G_{A}^{\lambda\mu\nu} \right]\,.
\end{equation}
Both bodies are tidally interacting (without dissipation), which is parametrized by a set of mass-type and current-type tidal polarizations $\{\mu_A^{(\ell)},\sigma_A^{(\ell)}\}$. These constants vanish for BHs~\cite{Damour:2009vw,Binnington:2009bb,Landry:2015zfa,LeTiec:2020bos}, but they are expected to be non-zero for neutron stars or exotic compact objects~\cite{Hinderer:2007mb,Cardoso:2017cfl}. They are linked to the dimensionless Love numbers $k^{(\ell)}$ and $j^{(\ell)}$ through
\begin{equation}\label{eq:muAlsigmaAl}
G \mu_A^{(\ell)} = \frac{2}{(2\ell-1)!!} k_A^{(\ell)}R_A^{2\ell +1}\,, \qquad G \sigma_A^{(\ell)} = \frac{\ell-1}{4(\ell+2)(2\ell-1)!!} j_A^{(\ell)}R_A^{2\ell +1}\,,
\end{equation}
where $R_A$ is the radius of body $A$. Tidal effects are seen a perturbation of the point-particle case. Hence, we use the tidal polarizabilities as a perturbing parameter in addition to the usual PN expansion. More precisely they are of order
\begin{equation}
\mu_A^{(2)} \sim \sigma_A^{(2)} = \calO (\etidal)\,, \qquad \mu_A^{(3)} = \calO \left(\frac{\etidal}{c^4}\right)\,,
\end{equation}
where $\etidal$ can be seen as a 5PN quantity. Thus, we will remain linear in $\etidal$ throughout this work. We also define the following convenient combinations of the tidal polarizabilities
\begin{equation}\label{eq:polarpm}
\mu_\pm^{(\ell)} = \frac{1}{2}\left(\frac{m_{2}}{m_{1}}\,\mu_{1}^{(\ell)} \pm
  \frac{m_{1}}{m_{2}}\,\mu_{2}^{(\ell)}\right)\,,\qquad \sigma_\pm^{(\ell)} =
\frac{1}{2}\left(\frac{m_{2}}{m_{1}}\,\sigma_{1}^{(\ell)} \pm
  \frac{m_{1}}{m_{2}}\,\sigma_{2}^{(\ell)}\right)\,,
\end{equation}
as well as their normalized version
\begin{equation}\label{eq:musigmatilde}
\widetilde{\mu}_\pm^{(\ell)} = \left(\frac{c^2}{G \tmass}\right)^{2\ell+1}
\!\!\!G\,\mu_\pm^{(\ell)}\,,\qquad \widetilde{\sigma}_\pm^{(\ell)} =
\left(\frac{c^2}{G \tmass}\right)^{2\ell+1} \!\!\!G\,\sigma_\pm^{(\ell)}\,.
\end{equation}
Most of the computations were done using the \textit{xTensor} extension~\cite{xtensor} of the \textit{Mathematica} software. 

\subsection{Spherical harmonics decomposition of the gravitational field}\label{subsec:sphericalharmonics}

The transverse-traceless (TT) projection $h_{ij}^\text{TT}$ of the gravitational field of an isolated matter system can be uniquely decomposed in terms of a set of STF mass and current multipole moments $\dU_L$ and $\dV_L$, called the radiative multipole moments, as \cite{Thorne:1980ru}
\begin{align}\label{eq:hij}
h_{ij}^\text{TT} &= \frac{4G}{c^2R} \,\mathcal{P}_{ijkl} (\bm{N}) \sum^\infty_{\ell=2}\frac{1}{c^\ell \ell !} \bigg[ N_{L-2} \,\dU_{klL-2}(T_R) - \frac{2\ell}{c(\ell+1)} \,N_{aL-2} \,\epsilon_{ab(k} \,\dV_{l)bL-2}(T_R)\bigg] +\mathcal{O}\left( \frac{1}{R^2} \right) \,,
\end{align}
where $R$ is the distance between the source and the observer, $\bm{N}$ is the direction of propagation of the GW, $\mathcal{P}_{ijkl}$ is the TT projector and $T_R= T- R/c$ is the retarded time in some radiative gauge in which $T_R$ is asymptotically null. From~\eqref{eq:hij}, one can derive the energy and angular momentum fluxes as functions of the radiative multipole moments
\begin{subequations}\label{eq:fluxes_moments}
\begin{align}
\mathcal{F} &= \sum_{\ell= 2}^{\infty} \frac{G}{c^{2\ell+1}} \frac{(\ell+1)(\ell+2)}{(\ell-1)\ell\,\ell !(2\ell+1)!!}\bigg[\dU_L^{(1)}\dU_L^{(1)} + \frac{4\ell^2}{c^2(\ell+1)^2}\,\dV_L^{(1)}\dV_L^{(1)}\bigg]\,,\\
\mathcal{G}_i &= \varepsilon_{iab}\,\sum_{\ell= 2}^{\infty} \frac{G}{c^{2\ell+1}} \frac{(\ell+1)(\ell+2)}{(\ell-1)\,\ell !(2\ell+1)!!}\bigg[\dU^{}_{aL-1}\dU_{bL-1}^{(1)} + \frac{4\ell^2}{c^2(\ell+1)^2}\,\dV^{}_{aL-1}\dV_{bL-1}^{(1)}\bigg]\,,
\end{align}
\end{subequations}
where the upper index $(\ell)$ refers to the $\ell^\text{th}$ time derivative. Next, the waveform polarizations are defined as
\begin{equation}\label{eq:hpluscross}
h_+ = \frac{1}{2}\bigl(P_i P_j - Q_i Q_j\bigr)h_{ij}^\text{TT}\,, \qquad
h_\times = \frac{1}{2}\bigl(P_i Q_j + Q_i P_j\bigr)h_{ij}^\text{TT}\,,
\end{equation}
where the vectors $(\bm{P},\bm{Q},\bm{N})$ form an orthonormal triad properly defined in \textit{e.g.} Sec. II. A. of~\cite{Henry:2022dzx}. As usual, we decompose $h_+ -\di h_\times$ in a spin-weighted spherical harmonics basis of weight -2 \cite{Kidder:2007rt}
\begin{equation}\label{eq:h}
h\equiv h_+ -\di h_\times = \sum_{\ell=2}^\infty \sum_{m=-\ell}^{\ell} h_{\ell m} Y^{\ell m}_{-2}(\Theta,\Phi),
\end{equation}
where the two angles $(\Theta,\Phi)$ characterize the direction of propagation $\bm{N}$. The gravitational modes are linked to the radiative moments by the relation~\cite{Faye:2012we}
\begin{equation}\label{eq:hlm}
h_{\ell m} = -\frac{2 G}{R c^{\ell+2}\ell !}\sqrt{\frac{(\ell+1)(\ell+2)}{\ell(\ell-1)}}\,\alpha_L^{\ell m} \left( \text{U}_L+\frac{2\ell}{\ell+1}\frac{\di}{c} \text{V}_L \right).
\end{equation}
Introducing a fixed orthonormal basis $(\boldsymbol{n}_0,\boldsymbol{\lambda}_0,\boldsymbol{l}_0)$ where $\boldsymbol{l}_0$ is the constant vector perpendicular to the orbital plane, together with $\boldsymbol{\mathfrak{m}}_0 = (\boldsymbol{n}_0+\di \boldsymbol{\lambda}_0)/\sqrt{2}$, the projector $\alpha_L^{\ell m}$ is explicitly given for positive $m$ by
\begin{equation}\label{eq:alphalm}
\alpha_L^{\ell m} = \frac{\sqrt{4\pi} (-\sqrt{2})^m\ell !}{\sqrt{(2\ell+1) (\ell+m)!\,(\ell-m)!}}\,\overline{\mathfrak{m}}_0^{\langle M}l_0^{L-M\rangle}\,,
\end{equation}
where the overbar denotes complex conjugation. To derive the full waveform amplitude to 2.5PN order, one needs to compute al the modes $h_{\ell m}$ for $\ell \leq 7$ and $\lvert m \rvert \leq 7$.

\subsection{Radiative in terms of source multipole moments}\label{subsec:genmoments}

The post-Newtonian-multipolar-post-Minkowskian (PN-MPM) formalism, see the living review~\cite{Blanchet:2013haa}, is valid for any compact-supported matter action. It allows to express the radiative multipole moments, defined in Eq.~\eqref{eq:hij}, to the so-called canonical multipole moments $\{\dM_L,\dS_L\}$. 
These canonical moments are further related to the source multipole moments $\{ \dI_L,\dJ_L\}$ and the gauge ones $\{\dW_L,\text{X}_L,\text{Y}_L,\text{Z}_L \}$. The up-to-date relations between these moments can be found in \textit{e.g.} Refs.~\cite{Faye:2014fra,Blanchet:2023bwj}. In this section, we detail all the relations that are required to derive the full waveform amplitude to 2.5PN order. One can split the mass and current radiative moments into several pieces, as 
\begin{subequations}\label{radtosource}
\begin{align}
\dU_L &= \dM_L^{(\ell)} + \dU_L^\text{tail} + \dU_L^\text{inst} + \dU_L^\text{mem}\,,\\
\dV_L &= \dS_L^{(\ell)} + \dV_L^\text{tail} + \dV_L^\text{inst}\,,
\end{align}
\end{subequations}
First, the tail part of the radiative moments is known for any $\ell$~\cite{Blanchet:1992br,Blanchet:1995fr} and reads
\begin{subequations}\label{eq:tails}
\begin{align}
    \dU_L^\text{tail} &= \frac{2G\ADM}{c^3} \int_0^\infty \dd \tau\left[ \ln\left( \frac{\tau}{2 b_0} \right) + \kappa_\ell \right] \dM_L^{(\ell+2)}(T_R-\tau)\,, \\
    \dV_L^\text{tail} &= \frac{2G\ADM}{c^3} \int_0^\infty \dd \tau\left[ \ln\left( \frac{\tau}{2 b_0} \right) + \pi_\ell \right] \dS_L^{(\ell+2)}(T_R-\tau)\,,
    \end{align}
\end{subequations}
where $\ADM$ refers to the Arnowitt-Deser-Misner (ADM) mass, $\{\kappa_\ell,\pi_\ell\}$ are constants of $\ell$ and $b_0$ is a gauge constant. Next, the instantaneous pieces that contribute to the 2.5PN waveform amplitude are given by
\begin{subequations}\label{eq:UVinstmom}
\begin{align}
\dU_{ij}^\text{inst} &= \frac{G}{7c^5} \biggl[ \dM_{a\langle i}^{(5)}\dM_{j\rangle a}^{\textcolor{white}{()}}-5 \dM_{a\langle i}^{(4)}\dM_{j\rangle a}^{(1)}-2 \dM_{a\langle i}^{(3)}\dM_{j\rangle a}^{(2)} +\frac{7}{3} \epsilon_{ab\langle i}\dM_{j\rangle a}^{(4)}\dS_b^{\textcolor{white}{()}}\biggr]+ \calO\left( \frac{1}{c^7} \right)\,,\\
\dU_{ijkl}^\text{inst} &= -\frac{G}{5c^3}\biggl[ 21 \dM_{\langle ij}^{(5)}\dM_{kl\rangle}^{\textcolor{white}{()}}+63 \dM_{\langle ij}^{(4)}\dM_{kl\rangle}^{(1)}+102 \dM_{\langle ij}^{(3)}\dM_{kl\rangle}^{(2)} \biggr]+ \calO\left( \frac{1}{c^5} \right)\,,\\
\dV_{ijk}^\text{inst} &= \frac{G}{10c^3}\biggl[ \epsilon_{ab\langle i}\left( \dM_{ja}^{(5)}\dM_{k\rangle b} -5 \dM_{ja}^{(4)}\dM_{k\rangle b}^{(1)}\right)-20 \,\dM^{(4)}_{\langle ij}\dS_{k\rangle}^{\textcolor{white}{(0)}} \!\!\biggr] + \calO\left( \frac{1}{c^5} \right)\,.
\end{align}
\end{subequations}
The instantaneous pieces of the other multipole moments do not contribute to the 2.5PN amplitude. Finally, the memory part (also called non-linear memory), only concerns the mass-type multipoles~\cite{Blanchet:1997jj}. For completeness, we write those contributing to the 2.5PN amplitude, although we left the memory computations for future works
\begin{subequations}\label{eq:Umem}
\begin{align}
\dU_{ij}^\text{mem} &= -\frac{2G}{7c^5}\int_{0}^{\infty}\dd \tau \, \dM_{a\langle i}^{(3)}(T_R-\tau)\dM_{j\rangle a}^{(3)}(T_R-\tau) +\calO\left( \frac{1}{c^7} \right)\,,\label{eq:Uijmem}\\
\dU_{ijkl}^\text{mem} &=\frac{2G}{5c^3}\int_{0}^{\infty}\dd \tau\, \dM_{\langle ij}^{(3)}(T_R-\tau)\dM_{kl\rangle}^{(3)}(T_R-\tau) +\calO\left( \frac{1}{c^5} \right)\,.
\end{align}
\end{subequations}
Finally, we only need to relate the canonical moments to the source (and gauge) ones. At 2.5PN, only the mass quadrupole contains a correction between the canonical and the source moment, it reads~\cite{Blanchet:1996wx}
\begin{equation}\label{eq:MijIij}
\dM_{ij} = \dI_{ij} + \frac{4G}{c^5}\left[ \dW^{(2)} \dI_{ij} - \dW^{(1)}\dI_{ij}^{(1)} \right] + \calO\left(\frac{1}{c^7}\right)\,.
\end{equation}
The other canonical moments can be replaced by the source multipole moments that we derived consistently in~\cite{Dones:2024odv}.

\subsection{The conservative and radiative dynamics for an eccentric motion}

In Paper I~\cite{paperI}, we solved the equations of motion at relative 2.5PN order. To do so, we first derived the conservative motion using a quasi-Keplerian parametrization (QKP) at NNLO with the conserved energy and angular momentum for starting point. In the presence of tidal effects, it takes the form
\begin{subequations}\label{eq:QKgen}
\begin{align}
r &= a_r(1-e_r \cos u)\,, \label{eq:QKr}\\
l = n(t-t_0) &= u - e_t \sin u + f_{v-u} (v-u) + \sum_{k=1}^{6}f_{kv} \sin(k v)\,,\label{eq:QKl}\\
\frac{\phi-\phi_0}{K} &= v+ \sum_{k=2}^{8}g_{kv} \sin(kv)\,,\label{eq:QKphi}\\
v&= 2 \arctan\left[\sqrt{\frac{1+e_\phi}{1-e_\phi}} \tan \frac{u}{2} \right]\,, \label{eq:QKv}
\end{align}
\end{subequations}
where $u$ is the eccentric anomaly, $l$ the mean anomaly, $a_r$ the semi-major axis, $n$ the mean motion and $K$ the periastron advance. All the coefficients, especially the different eccentricities, contain tidal corrections and their expressions in terms of the conserved quantities are displayed in Appendix~B of Paper~I. Next, we derived the separation, phase and their time derivatives $(r,\rd,\phi,\fid)$ in terms of the eccentric anomaly, the orbital phase and the time eccentricity $(x,\et,u)$ where
\begin{equation}
x= \left( \frac{G \tmass \Omega}{c^3}\right)^{2/3},
\end{equation}
with $\Omega = K n$ being the orbital frequency. Then, we inverted the generalized Kepler equation to obtain their values in terms of $(x,\et,l)$ at the cost of a small-eccentricity expansion up to $\calO\bigl(\et^{14}\bigr)$. This is due to the fact that our goal is to provide the waveform at $\calO(\et^{12})$ but some division by the eccentricity in the intermediate results forces one to compute higher orders in the eccentricity expansions. We recall that the eccentricity expanded quantities cannot be taken above the maximum eccentricity $e_\text{max}\simeq 0.6627434$, see Section~III.~C.~3. of Paper~I. With this inversion formula at hand, we were able to obtain the expressions of $(r,\rd,\fid)$ in terms of $(x,\et,l)$, while the phase is split into a secular part $\lambda = K l$ and an oscillating part $W(l)$
\begin{equation}\label{eq:phil}
\phi(x,\et,l) -\phi_0 = \lambda + W(l)\,.
\end{equation}

Then, we turned to the so-called post-adiabatic (PA) corrections, which are necessary for the computation of the (2,2) and (2,0) modes. We included the effects of the radiation from the equations of motion derived in~\cite{Dones:2024odv} using a method of variations of the constants developed in~\cite{Damour:2004bz,Konigsdorffer:2006zt,Boetzel:2019nfw}. We have generalized this method to the particular QKP~\eqref{eq:QKgen} to deal with tidal effects at leading PA order. The idea is to allow for a time-dependency on four variables that we chose to be $\{x,\et,l,\lambda\}$. Each quantity expressed in terms of those variables can be split in secularly evolving and rapidly oscillating parts. The secular parts are denoted with a bar, and the oscillating parts with a tilde, for example: $x(t)=\bar{x}(t)+\tilde{x}(t)$.

The computations of the present paper are based upon the expressions of the source multipole moments derived in~\cite{Dones:2024odv}, valid for arbitrary motion, and the solution of the equations of motion at relative 2.5PN computed in~Paper~I~\cite{paperI}. This allows to derive the energy and angular momentum fluxes from which we can deduce the secular evolution of the orbital elements (useful notably for Phenom models) and the amplitude of the GW strain, decomposed in spin-weighted spherical harmonics. 

\section{Radiated fluxes and secular evolution of the orbital elements}\label{sec:fluxes}

To compute the fluxes at relative 2.5PN, we truncate the sums~\eqref{eq:fluxes_moments} to $\ell=4$ because the other radiative multipoles contribute to higher orders. This means that we will deal with the two following expressions
\begin{subequations}\label{eq:fluxesRad2PN}
\begin{align}
\mathcal{F} &= \dfrac{G}{c^{5}}\left\{ \dfrac{1}{5}\dU_{ij}^{(1)}\dU_{ij}^{(1)} + \dfrac{1}{c^{2}} \left[ \dfrac{1}{189} \dU_{ijk}^{(1)}\dU_{ijk}^{(1)} + \dfrac{16}{45}\dV_{ij}^{(1)}\dV_{ij}^{(1)}  \right] + \dfrac{1}{c^{4}} \left[ \dfrac{1}{9072}\dU_{ijkm}^{(1)}\dI_{ijkm}^{(1)} + \dfrac{1}{84}\dV_{ijk}^{(1)}\dV_{ijk}^{(1)} \right] 
\right\}\,,\\
\mathcal{G}^i &= \dfrac{G}{c^{5}}\epsilon_{iab}\left\{ \dfrac{2}{5}\dU_{ak}\dU_{bk}^{(1)} + \dfrac{1}{c^{2}} \left[ \dfrac{1}{63} \dU_{akl}\dU_{bkl}^{(1)} + \dfrac{32}{45}\dV_{ak}\dV_{bk}^{(1)}  \right] + \dfrac{1}{c^{4}} \left[ \dfrac{1}{2268}\dU_{aklm}\dU_{bklm}^{(1)} + \dfrac{1}{28}\dV_{akl}\dV_{bkl}^{(1)} \right] 
\right\}\,.
\end{align}
\end{subequations}
The instantaneous, tail and memory contributions to the radiated fluxes can be computed separately. Notice that the memory terms~\eqref{eq:Umem} become instantaneous contributions when time-differentiated. Hence, there is no memory contribution to the energy flux, but in principle the angular momentum flux contains one which is not necessarily 0 for eccentric orbits. This contribution has been computed only for point-particles at LO, first in~\cite{Arun:2009mc}, and then corrected in Ref.~\cite{Trestini:2024mfs} which showed that it vanishes after orbit-averaging. Here, the computation of memory contributions to the angular momentum flux for higher PN orders or tidal effects are left for future works. In~\Cref{subsec:fluxinst}, we compute the orbit-averaged instantaneous fluxes, in~\Cref{subsec:fluxtail}, we compute the orbit-averaged tail fluxes together with an eccentricity resummation and in~\Cref{subsec:secularOrb}, we use the total fluxes to derive the differential equations describing the secular evolutions of the orbital elements to relative 2.5PN.

\subsection{Instantaneous part of the radiated fluxes}\label{subsec:fluxinst}
\subsubsection{Generic orbits}

We start from the expressions of the fluxes~\eqref{eq:fluxesRad2PN} expressed in terms of the radiative multipole moments. We neglect the tail contributions and write the fluxes in terms of the source and gauge multipole moments using~\eqref{radtosource},~\eqref{eq:UVinstmom} and~\eqref{eq:MijIij}. Then, we use the expressions of the source moments derived in~\cite{Dones:2024odv}, and compute consistently their time derivatives using the relative acceleration at 2.5PN. This leads to the instantaneous part of the radiative moments, expressed in terms of $(r,\rd,\phi,\fid)$. Finally, we deduce the fluxes in terms of $(r,\rd,\fid)$ which read at LO
\begin{subequations}\label{eq:FGinstgen}
\begin{align}
\mathcal{F}_\text{inst} &=\frac{32}{5}\frac{G^3\tmass^4\nu^2}{r^4c^5}\Biggl\{ \bigl(r\fid\bigr)^2 +\frac{\rd^2}{12} + \frac{\mu_+^{(2)} + \delta \, \mu_-^{(2)}}{\tmass\nu r^4} \left[ 15 \rd^4 -\frac{225}{2}\bigl(\rd r \fid\bigr)^2 + \frac{45}{2} \bigl(r\fid\bigr)^4 + \frac{G\tmass}{2 r} \left( 21 \rd^2 -33 \bigl( r\fid \bigr)^2 \right) \right]\nn \\
& \qquad \qquad \qquad \quad + \frac{12 G \mu_+^{(2)}}{r^5} \Bigl( 3 \bigl(r\fid\bigr)^2 -\rd^2 \Bigr) \Biggr\}\,,\\
\mathcal{G}^i_\text{inst} &=\frac{16}{5}\frac{G^2\tmass^3\nu^2\fid}{rc^5}\Biggl\{ \bigl(r\fid\bigr)^2 - \frac{\rd^2}{2} +\frac{G\tmass}{r} + \frac{\mu_+^{(2)} + \delta \, \mu_-^{(2)}}{\tmass\nu r^4} \left[ 90 \rd^4 -135\bigl(\rd r \fid\bigr)^2 + \frac{45}{4} \bigl(r\fid\bigr)^4 + \frac{G\tmass}{r} \left( \frac{27}{2}\bigl( r\fid \bigr)^2 -\frac{117}{4} \rd^2 \right)\right.\nn \\
& \qquad \qquad \qquad \qquad \quad \left. -\frac{51}{4} \frac{G^2\tmass^2}{r^2}\right] + \frac{G \mu_+^{(2)}}{r^5} \left( 18 \bigl(r\fid\bigr)^2 -54 \rd^2+36 \frac{G\tmass}{r} \right) \Biggr\}\ell^i \,.
\end{align}
\end{subequations}
The full expressions at relative 2.5PN are provided in the ancillary files~\cite{SuppMaterial2}.

\subsubsection{Orbit averaged on eccentric orbits}

Starting from the instantaneous radiated fluxes~\eqref{eq:FGinstgen} at relative 2.5PN, we replace $(r,\rd,\fid)$ in terms of $(x,\et,u)$ derived in Section~III.~B. of Paper~I. Next, we compute their orbit average, where for a given $P$-periodic function $A$, is given by the following integral
\begin{equation}
\langle A\rangle = \frac{1}{P} \int_0^P \dd t \, A = \int_0^{2\pi}\frac{\dd u}{2\pi} \frac{\dd l}{\dd u} A(u)\,,
\end{equation}
where we recall that $l(u)$ is given in~\eqref{eq:QKl} and the explicit expression of $\dd l/\dd u$ can be found in Eq.~(3.19) of Paper~I. Notice that at the 2.5PN order, one should in principle take into account the PA corrections. However, it is not necessary here since they appear at an odd PN order and it is known that instantaneous terms at odd PN orders in the fluxes vanish when orbit averaged because they are odd functions of $u$ or $l$. In the absence of logarithms which appear at 3PN, we need to evaluate the following kernel integrals, for $n\geq 1$,
\begin{subequations}
\begin{align}
\int_0^{2\pi}\frac{\dd u}{2\pi} \frac{\cos{(ku)}}{(1-e\cos{u})^n}&=  \frac{(n+k-1)!}{(n-1)!}\beta^k\sum_{\ell=0}^{n-1}\frac{1}{2^\ell \ell! (k+\ell)!}\frac{(n+\ell-1)!}{ (n-\ell-1)!}\frac{(1-\sqrt{1-e^2})^{\ell}}{(1-e^2)^{(n+\ell)/2}}\,,\label{eq:Ikn}\\
\int_0^{2\pi}\frac{\dd u}{2\pi} \frac{\sin{(ku)}}{(1-e\cos{u})^n}&= 0\,,
\end{align}
\end{subequations}
where $\beta = \tfrac{1-\sqrt{1-e^2}}{e}$. In Appendix~\ref{app:Ikn}, we provide a proof of this expression, and we extend the computation to the integrals containing logarithms. By combining the fluxes computed in the previous Section with the expression of $\dd l/\dd u$ in terms of $(x,\et,u)$, we get the orbit-averaged energy flux at NNLO. Here, we only display its value at LO exact in eccentricity
\begin{align}\label{eq:FinstOrbAvg}
\langle \mathcal{F}_\text{inst}  \rangle =& \frac{32x^5\nu^2  c^5}{5G(1-\et^2)^{7/2}}\left(1+\frac{73}{24} \et^2+\frac{37}{96}\et^4\right)\nn\\
& + \frac{192c^5\nu x^{10}}{5G(1-\et^2)^{17/2}} \Biggl\{\Bigl(\mutp+ \delta\,\mutm \Bigr)\left[ 1 + \frac{211}{8}\et^2+\frac{3369}{32}\et^4+\frac{6275}{64}\et^6+\frac{10355}{512}\et^8 + \frac{225}{512}\et^{10}\right] \nn \\
&  \qquad \qquad \qquad \qquad + \nu \, \mutp \left[-3 + \frac{1247}{12}\et^2+\frac{56069}{192}\et^4+\frac{5341}{32}\et^6+\frac{42019}{3072}\et^8 \right. \nn\\
& \qquad \qquad \qquad \qquad \qquad \qquad \left. + \sqrt{1-\et^2}\left( 7 + \frac{1327}{24}\et^2+\frac{1081}{24}\et^4+\frac{3335}{384}\et^6+\frac{37}{192}\et^8\right) \right] \Biggr\}\,.
\end{align}
Since the motion is planar, the unitary vector $\ell^i$, which is along the direction of the orbital angular momentum, is constant and matches $\bm{l}_0$ defined in~\Cref{subsec:sphericalharmonics}. Hence, we only consider the norm of the angular momentum flux and drop the index $i$. It reads
\begin{align}\label{eq:GinstOrbAvg}
\langle \mathcal{G}_\text{inst}  \rangle &= \frac{32 \tmass c^2 \nu^2 x^{7/2}}{5(1-\et^2)^2}\left(1+\frac{7}{8} \et^2\right)\nn\\
& + \frac{192\tmass c^2\nu x^{17/2}}{5(1-\et^2)^7} \Biggl\{\Bigl( \mutp+\delta\,\mutm\Bigr)\left[ 1 + \frac{117}{8}\et^2+\frac{915}{32}\et^4+\frac{635}{64}\et^6+\frac{165}{512}\et^8 \right]  \\
& \qquad \qquad \qquad \qquad + \nu \,\mutp \left[ \frac{251}{4}\et^2+\frac{5625}{64}\et^4+\frac{143}{8}\et^6+\frac{15}{64}\et^8 + \sqrt{1-\et^2}\left( 4 + \frac{71}{4}\et^2+\frac{95}{16}\et^4+\frac{7}{16}\et^6\right)\right]\Biggr\}\nn \,.
\end{align}
These expressions are exact in eccentricity. The point-particle part of these instantaneous fluxes, derived in harmonic coordinates, are in agreement with those of~\cite{Arun:2007sg,Arun:2009mc} up to 2PN.

\subsection{Tail part of the radiated fluxes}\label{subsec:fluxtail}

\subsubsection{Orbit averaged}

The tail part of the radiative moments, explicited in~\eqref{eq:tails}, requires the knowledge of the ADM mass at NLO
\begin{equation}\label{eq:mADM}
\ADM = \tmass\left[ 1-\frac{\nu x}{2} -\frac{3}{2} \nu \mutp x^6 \left( \frac{4+\et^2}{(1-\et^2)^{7/2}} - \frac{10+15\et^2+\frac{5}{4}\et^4}{(1-\et^2)^5} \right) \right] + \calO\left(\frac{1}{c^4},\frac{\etidal}{c^4}\right)\,.
\end{equation}
Since we perform the computation at relative 2.5PN, we need to compute the NLO tail of the mass quadrupole~$\dU_{ij}^\text{tail}$ and the LO contributions of the current quadrupole and mass octupole, $\dV_{ij}^\text{tail}$ and $\dU_{ijk}^\text{tail}$. When combining all the information at the relevant order, the tail part of the fluxes are obtained by computing the following time integrals
\begin{subequations}
\begin{align}
\mathcal{F}_\text{tail} =& \,\frac{4G^2\ADM}{c^8}\Biggl\{\frac{1}{5} \dI_{ij}^{(3)}(T_R)\int_0^\infty\dd \tau \ln\left( \frac{\tau}{\tau_1} \right)\dI_{ij}^{(5)}(T_R-\tau) \nn \\
&+ \frac{1}{c^2}\left[\frac{16}{45} \dJ_{ij}^{(3)}(T_R)\int_0^\infty\dd \tau \ln\left( \frac{\tau}{\tau_2} \right)\dJ_{ij}^{(5)}(T_R-\tau) + \frac{1}{189} \dI_{ijk}^{(4)}(T_R)\int_0^\infty\dd \tau \ln\left( \frac{\tau}{\tau_3} \right)\dI_{ijk}^{(6)}(T_R-\tau) \right]\Biggr\}\,,\\
\mathcal{G}^i_\text{tail} =& \,\frac{G^2\ADM}{c^8} \epsilon_{iab}\Biggl\{\frac{4}{5}\left[ \dI_{ak}^{(2)}(T_R)\int_0^\infty\dd \tau \ln\left( \frac{\tau}{\tau_1} \right)\dI_{bk}^{(5)}(T_R-\tau) - \dI_{ak}^{(3)}(T_R)\int_0^\infty\dd \tau \ln\left( \frac{\tau}{\tau_1} \right)\dI_{bk}^{(4)}(T_R-\tau)\right] \nn \\
& + \frac{64}{45c^2}\left[ \dJ_{ak}^{(2)}(T_R)\int_0^\infty\dd \tau \ln\left( \frac{\tau}{\tau_2} \right)\dJ_{bk}^{(5)}(T_R-\tau) - \dJ_{ak}^{(3)}(T_R)\int_0^\infty\dd \tau \ln\left( \frac{\tau}{\tau_2} \right)\dJ_{bk}^{(4)}(T_R-\tau)\right] \nn \\
& + \frac{2}{63c^2}\left[ \dI_{akm}^{(3)}(T_R)\int_0^\infty\dd \tau \ln\left( \frac{\tau}{\tau_3} \right)\dI_{bkm}^{(6)}(T_R-\tau) - \dI_{akm}^{(4)}(T_R)\int_0^\infty\dd \tau \ln\left( \frac{\tau}{\tau_3} \right)\dI_{bkm}^{(5)}(T_R-\tau)\right] \Biggr\}\,,
\end{align}
\end{subequations}
with $\tau_1 = 2b_0 e^{-11/12}$, $\tau_2 = 2b_0 e^{-7/6}$ and $\tau_3 = 2b_0 e^{-97/60}$. 
In order to integrate these, we use the source multipole moments computed in~\cite{HFB20b,Dones:2024odv} and differentiate them with respect to time. This leads to expressions in terms of $(r,\rd,\phi,\fid)$ which we substitute by expressing it in terms of $(x,\et,l,\lambda)$, where the phase has been split into its secular and oscillating parts using~\eqref{eq:phil}. Finally, in order to perform the time integration, we need to evaluate integrals of the form
\begin{equation}\label{eq:tailkernel}
\forall (k,\Omega)\in\mathbb{N}\times\mathbb{R}^*, \quad \int_0^\infty \dd\tau \, \tau^k \ln\left(\frac{\tau}{p}\right) \e^{\di\, \Omega\tau} = \frac{\di^k k!}{\Omega^{k+1}}\Biggl[ -\frac{\pi}{2}\text{sign}(\Omega) -\di\Bigl[\ln\bigl(p|\Omega|\bigr) + \gamma_E - H_k \Bigr]   \Biggr]\,,    
\end{equation}
where $\gamma_E$ is the Euler constant and $H_k$ is the $k^\text{th}$ harmonic number and $p>0$. Note that it is possible to compute the tail part of the fluxes without performing an eccentricity expansion using infinite sums of Bessel functions, see \textit{e.g.}~\cite{Arun:2007rg}, which define a set of so-called enhancement functions that require to be numerically integrated. For practical reasons, we did not perform similar computations in the present paper. Instead, we make the choice of truncating in eccentricity the integrands, which does not require to numerically evaluate the enhancement functions. We chose to perform the eccentricity expansion up to the fourteenth order, this order is motivated by Phenom waveform models accuracy, see discussion in~\Cref{sec:modes}. The LO results of the eccentricity-expanded tail fluxes at $\calO(\et^{14})$ after orbit averaging read
\begin{subequations}
\begin{align}
\label{eq:Ftailexp}\langle\mathcal{F}_\text{tail}\rangle =& \,\frac{128\pi x^{13/2}\nu^2 c^5}{5G} \left[1 +\frac{2335}{192}\et^2 +\frac{42955}{768}\et^4 + \frac{6204647}{36864} \et^6 +\frac{352891481}{884736}\et^8 \right] \nn\\
& + \frac{768\pi x^{23/2}\nu c^5}{5G}\biggl[ \Bigl(\mutp+ \delta\,\mutm \Bigr) \left( 1 +\frac{7015}{128}\et^2 +\frac{52655}{64}\et^4 + \frac{478158179}{73728} \et^6 +\frac{10183919287}{294912}\et^8 \right) \nn \\
& \qquad \qquad \qquad \qquad  + \nu\, \mutp\left( 4 +\frac{118255}{384}\et^2 +\frac{1490735}{384}\et^4 + \frac{648156127}{24576} \et^6 +\frac{13865426459}{110592}\et^8  \right) \biggr] 
\,,\\
\label{eq:Gtailexp}\langle\mathcal{G}_\text{tail}\rangle =& \, \frac{128\pi x^5\tmass \nu^2 c^2}{5} \left[1 +\frac{209}{32}\et^2 +\frac{2415}{128}\et^4 + \frac{730751}{18432} \et^6 +\frac{10355719}{147456}\et^8 \right]\nn\\
& + \frac{768\pi x^{10}\tmass\nu c^2}{5}\biggl[ \Bigl(\mutp+ \delta\,\mutm \Bigr) \left( 1 +\frac{2233}{64}\et^2 +\frac{23695}{64}\et^4 + \frac{81191887}{36864} \et^6 +\frac{1364684119}{147456}\et^8 \right) \nn \\
& \qquad \qquad \qquad \qquad  + \nu\, \mutp\left( 4 +\frac{6081}{32}\et^2 +\frac{460265}{256}\et^4 + \frac{88916029}{9216} \et^6 +\frac{3680888209}{98304}\et^8 \right) \biggr] 
\,,
\end{align}
\end{subequations}
where the orbit average has been obtained by setting all $e^{\di n l}$, $n\neq 0$, to 0 after the time integration.

\subsubsection{Resummed tail fluxes}\label{subsubsec:resum}

As was done in~\cite{Henry:2023tka}, we perform an eccentricity resummation on the tail part of the fluxes~\eqref{eq:Ftailexp}-\eqref{eq:Gtailexp}. To do so, we perform the ansatz that the power of $x$ is linked to the power of the prefactor $(1-\et^2)$ the same way as they are in the instantaneous part of the fluxes (see Eqs.~\eqref{eq:FGtot}). The resummed coefficients are obtained by matching the eccentricity expansion of the ansatz with the eccentricity expanded tail fluxes. We find at LO and $\calO(\et^{8})$
\begin{subequations}\label{eq:FGtailresum}
\begin{align}
\label{eq:Ftailresum}\langle\mathcal{F}^\text{resum}_\text{tail}\rangle =&\, \frac{128\pi x^{13/2}\nu^2 c^5}{5G(1-\et^2)^5} \left[1 +\frac{1375}{192}\et^2 +\frac{3935}{768}\et^4 + \frac{10007}{36864} \et^6 +\frac{2321}{884736}\et^8 +\calO\bigl(\et^{10}\bigr) \right]\nn\\
& + \frac{768\pi x^{23/2}\nu c^5}{5G(1-\et^2)^{10}}\left[ \Bigl(\mutp+ \delta\,\mutm \Bigr) \left( 1 +\frac{5735}{128}\et^2 +\frac{5115}{16}\et^4 + \frac{44554019}{73728} \et^6 +\frac{98557247}{294912}\et^8 \right) \right. \nn \\
& \qquad \qquad \qquad \qquad \left. + \nu\, \mutp\left( 4 +\frac{102895}{384}\et^2 +\frac{377305}{384}\et^4 + \frac{22863647}{24576} \et^6 +\frac{3041353}{13824}\et^8 \right) + \calO\bigl(\et^{10}\bigr)\right]\,,\\
\label{eq:Gtailresum}\langle\mathcal{G}^\text{resum}_\text{tail}\rangle =&\,\frac{128\pi x^5\tmass \nu^2 c^2}{5(1-\et^2)^{7/2}} \left[1 +\frac{97}{32}\et^2 +\frac{49}{128}\et^4 - \frac{49}{18432} \et^6 - \frac{109}{147456}\et^8  +\calO\bigl(\et^{10}\bigr)\right]\nn\\
& + \frac{768\pi x^{10}\tmass\nu c^2}{5(1-\et^2)^{17/2}}\left[ \Bigl(\mutp+ \delta\,\mutm \Bigr) \left( 1 +\frac{1689}{64}\et^2 +\frac{13509}{128}\et^4 + \frac{3633127}{36864} \et^6 +\frac{1002787}{49152}\et^8 \right) \right. \nn \\
& \qquad \qquad \qquad \qquad \left. + \nu\, \mutp\left( 4 +\frac{4993}{32}\et^2 +\frac{79397}{256}\et^4 + \frac{1352599}{9216} \et^6 +\frac{21139}{294912}\et^8 \right)+\calO\bigl(\et^{10}\bigr) \right] \,.
\end{align}
\end{subequations}
Note that this resummation is not unique since the leading order of the instantaneous fluxes contain a polynomial of the eccentricity and the product of $\sqrt{1-\et^2}$ with another polynomial. Of course, one could choose another resummation, see \textit{e.g.} the discussion in Section~IV.~D. of~\cite{Trestini:2025yyc}. However, as explained in Section~III.~D. of~\cite{Henry:2023tka}, we showed that this resummation at leading point-particle order has a satisfactory precision when compared to the numerically evaluated enhancement functions, with a relative error of $\sim 10^{-4}$ for eccentricities between 0 and 1. 

The total fluxes that we will use in the rest are the sum of the instantaneous orbit-averaged fluxes~\eqref{eq:FinstOrbAvg}-\eqref{eq:GinstOrbAvg} with the resumed tail fluxes~\eqref{eq:FGtailresum}, which take the form
\begin{subequations}\label{eq:FGtot}
\begin{align}
\langle \mathcal{F}\rangle =& \,\frac{32 x^5 c^5 \nu^2}{5G(1-\et^2)^{7/2}}\biggl[ \mathcal{F}_{0} + \frac{x}{1-\et^2} \mathcal{F}_{1} + \frac{\pi\,x^{3/2}}{(1-\et^2)^{3/2}} \mathcal{F}_{1.5} + \frac{x^2}{(1-\et^2)^2} \mathcal{F}_{2} + \frac{\pi\,x^{5/2}}{(1-\et^2)^{5/2}} \mathcal{F}_{2.5} \biggr] \nn \\
& + \frac{192 x^{10} c^5 \nu}{5G(1-\et^2)^{17/2}} \biggl[ \mathcal{F}_{5} + \frac{x}{1-\et^2} \mathcal{F}_{6} + \frac{\pi\,x^{3/2}}{(1-\et^2)^{3/2}} \mathcal{F}_{6.5} + \frac{x^2}{(1-\et^2)^2}\mathcal{F}_{7} + \frac{\pi\,x^{5/2}}{(1-\et^2)^{5/2}} \mathcal{F}_{7.5} \biggr]\,,\\
\langle \mathcal{G}\rangle =& \,\frac{32 \tmass c^2 \nu^2 x^{7/2}}{5(1-\et^2)^2}\biggl[ \mathcal{G}_{0} + \frac{x}{1-\et^2} \mathcal{G}_{1} + \frac{\pi\,x^{3/2}}{(1-\et^2)^{3/2}} \mathcal{G}_{1.5} + \frac{x^2}{(1-\et^2)^2} \mathcal{G}_{2} + \frac{\pi\,x^{5/2}}{(1-\et^2)^{5/2}} \mathcal{G}_{2.5} \biggr] \nn \\
& + \frac{192\tmass c^2\nu x^{17/2}}{5(1-\et^2)^7} \biggl[ \mathcal{G}_{5} + \frac{x}{1-\et^2} \mathcal{G}_{6} + \frac{\pi\,x^{3/2}}{(1-\et^2)^{3/2}} \mathcal{G}_{6.5} + \frac{x^2}{(1-\et^2)^2} \mathcal{G}_{7} + \frac{\pi\,x^{5/2}}{(1-\et^2)^{5/2}} \mathcal{G}_{7.5} \biggr]\,.
\end{align}
\end{subequations}
The different coefficients are listed in Appendix~\ref{app:fluxOrbAvg}. In the limit $\et\rightarrow 0$, we recover the energy flux computed in Eq.~(4.3) of Ref.~\cite{Dones:2024odv} which includes the adiabatic tides at the same PN order in the quasi-circular approximation. The angular momentum flux has not been derived in that reference since we know that it is related to the energy flux with the relation $\mathcal{F} = \Omega \,\mathcal{G}$ on circular orbits~\cite{Damour:1999cr,Blanchet:2001id,LeTiec:2011ab}.

\subsection{Secular evolution of orbital elements}\label{subsec:secularOrb}

For a given quantity $A$ depending on the conserved quantities of the problem, its secular evolution is computed using the fluxes balance equations $\langle \mathcal{F} \rangle = - \langle \tfrac{\dd E}{\dd t} \rangle$ and $\langle \mathcal{G} \rangle = - \langle \tfrac{\dd J}{\dd t} \rangle$ through
\begin{equation}
\langle \dot{A} \rangle = - \frac{\partial A}{\partial E} \langle \mathcal{F} \rangle -\frac{\partial A}{\partial J} \langle \mathcal{G} \rangle\,,
\end{equation}
%
where $J = |\bm{J}|$ is the norm of the conserved angular momentum. The expressions of the orbital elements in terms of the conserved quantities are available in Paper~I. This allowed to compute the secular evolution of $(x,\et,n,a_r,k)$ at relative 2.5PN, using the total orbit averaged resummed fluxes~\eqref{eq:FGtot}. We display here only the LO part of $\dot{x}$ and $\dot{e}_t$
\begin{subequations}\label{eq:xdotedot}
\begin{align}
\langle \dot{x} \rangle =& \frac{64 \, x^5\nu c^3}{5G\tmass(1-\et^2)^{7/2}}\left(1+\frac{73}{24} \et^2+\frac{37}{96}\et^4\right)\nn\\
& + \frac{384\, x^{10} c^3}{5G\tmass(1-\et^2)^{17/2}} \Biggl\{\Bigl(\mutp+ \delta\,\mutm \Bigr)\left[ 1 + \frac{211}{8}\et^2+\frac{3369}{32}\et^4+\frac{6275}{64}\et^6+\frac{10355}{512}\et^8 + \frac{225}{512}\et^{10}\right] \nn \\
&  \qquad \qquad \qquad \qquad \qquad + \nu \, \mutp \left[27 + \frac{549}{4}\et^2+\frac{7303}{24}\et^4+\frac{57727}{384}\et^6+\frac{39199}{3072}\et^8 \right. \nn\\
& \qquad \qquad \qquad \qquad \qquad \qquad \left. + \sqrt{1-\et^2}\left( -5 + \frac{1237}{24}\et^2+\frac{3869}{64}\et^4+\frac{1813}{192}\et^6 -\frac{29}{128}\et^8\right) \right]\Biggr\}\,,\\
\langle \dot{e}_t \rangle =& - \frac{304 \, x^4\nu c^3 \et}{15G\tmass(1-\et^2)^{5/2}}\left(1+\frac{121}{304} \et^2\right)\nn \\
& - \frac{2448\, x^9 c^3\et}{5G\tmass(1-\et^2)^{15/2}} \Biggl\{\Bigl(\mutp+ \delta\,\mutm \Bigr)\left[ 1 + \frac{487}{68}\et^2+\frac{1245}{136}\et^4+\frac{2545}{1088}\et^6+\frac{65}{1088}\et^8 \right] \nn \\
&  \qquad \qquad \qquad \qquad \qquad + \nu \, \mutp \left[\frac{479}{102} + \frac{27611}{1224}\et^2+\frac{96463}{4896}\et^4+\frac{100463}{39168}\et^6+\frac{5}{272}\et^8 \right. \nn\\
& \qquad \qquad \qquad \qquad \qquad \qquad \left. + \sqrt{1-\et^2}\left( \frac{550}{153} + \frac{7391}{1224}\et^2+\frac{4057}{4896}\et^4 -\frac{13}{288}\et^6 \right) \right]\Biggr\} \,.
\end{align}
\end{subequations}
These correspond exactly to Eqs.~(4.21) of Paper~I. However, with the present method, we are able to compute the secular corrections up to 2.5PN beyond LO while the method explained in Section~IV of Paper~I can only be consistent at LO (relative 0PN). The full expressions are provided in the ancillary file~\cite{SuppMaterial2}. In Appendix~\ref{subsec:SecExpr}, we display the LO expressions for $(n,k,a_r)$. As a consistency check, we verified that Eq.~(4.5) of~\cite{Dones:2024odv} is recovered by taking $\et=0$ in the relative 2.5PN expression of $\langle \dot{x} \rangle$. The family of Phenom models describe the dynamics by numerically integrating $\langle \dot{x} \rangle$, $\langle \dot{e}_t \rangle$, $\langle \dot{l} \rangle$ and $\langle \dot{\lambda} \rangle$, which is why we derived them to consistent high orders. We recall that~$\langle \dot{l} \rangle = \langle n(t)\rangle$ and $\langle \dot{\lambda} \rangle = \langle K(t) n(t)\rangle$, see Section~IV. of Paper~I.

The system~\eqref{eq:xdotedot} constitutes a coupled system of two differential equations of two variables. We leave its analytical resolution for future works, however let us comment on the procedure to follow. We cannot compute exact solutions of such system. The idea is to remove the time-dependency by solving the differential equation
\begin{equation}
\frac{\dd \bar{x}}{\dd \bar{e}_t} = \frac{\langle \dot{x} \rangle}{ \langle \dot{e}_t \rangle} = f(\bar{x},\bar{e}_t)\,,
\end{equation}
where the function $f$ is obtained by performing PN and eccentricity expansions. Next, one needs to come up with an ansatz on the solution $x(\et)$, introducing the initial orbital frequency $x_0$ and time eccentricity $e_0$ and inject it in the differential equation to fix the parameters of the ansatz. This procedure has been done in various works including 3PN point-particle~\cite{Moore:2016qxz,Boetzel:2019nfw} and spins~\cite{Henry:2023tka,Sridhar:2024zms}. This solution is required to derive the so-called ``DC" memory part of the amplitude modes, which are not dealt with in the present work. In~\Cref{sec:num}, we solve numerically the system~\eqref{eq:xdotedot}.

\section{Waveform amplitude}\label{sec:modes}

Previous works~\cite{Boetzel:2019nfw,Ebersold:2019kdc,Henry:2023tka} have provided eccentricity expanded expressions for the gravitational waveform modes up to order $\calO(\et^6)$. These expressions have been implemented in inspiral-merger-ringdown Phenom models~\cite{Planas:2025feq}, and showed to be accurate with respect to numerical relativity up to eccentricities of 0.3 defined at 20Hz~\cite{Planas:2025feq}, however, for eccentricities of 0.5 and above, the lack of higher orders in eccentricity causes unphysical features in the waveform due to the missing higher mean anomaly harmonic terms which are proportional to higher orders in eccentricity~\cite{Planas:2025jny}. Therefore, in order to overcome this limitation, we wish to provide eccentricity expanded expressions of the GW strain up to $\calO(\et^{12})$. To be consistent at this eccentricity order, it is necessary to compute some intermediate quantities at the next (non-vanishing) order due to the presence of some division by the eccentricity.

As shown in~\eqref{eq:h}, the GW strain is decomposed in spin-weighted spherical harmonics. Similarly to the fluxes, the modes can be split in three different effects: instantaneous, tail and memory, which we symbolically write
\begin{equation}
h_{\ell m} = h_{\ell m}^\text{inst} + h_{\ell m}^\text{tail} + h_{\ell m}^\text{mem} 
\,.
\end{equation}
The instantaneous part can be further split in two contributions: the ``adiabatic" and post-adiabatic parts, in which we take into account the radiation reaction to the dynamics computed in Paper~I. We recall that the memory contributions have been left for future work. In the following, due to the length of the results, we will display only the (2,2) mode at low PN  and eccentricity orders, however we recall that the ancillary file~\cite{SuppMaterial2} contains all modes from $\ell=2$ to $\ell=7$ to consistent relative 2.5PN order and eccentricity expanded up to $\calO\bigl(\et^{12}\bigr)$. We also define the convenient normalized mode $H_{\ell m}$, which is a function of only $(x,\et,l)$, as
\begin{equation}
h_{\ell m} = \frac{8 G \tmass \nu x}{R\, c^2}\sqrt{\frac{\pi}{5}} H_{\ell m}(x,\et,l) e^{-\di m \phi}\,.
\end{equation}
In~\Cref{subsec:modesinst}, we derive the instantaneous contributions; in~\Cref{subsec:modestail} the tail contributions; in~\Cref{subsec:modesPA} the PA corrections to the $(2,2)$ and $(2,0)$ modes; and finally in~\Cref{subsec:modespsi}, we compute the full waveform including the observable phase $\psi$ of the GW.

\subsection{Instantaneous part}\label{subsec:modesinst}

The procedure to derive the instantaneous part of the mode is sensibly identical to the derivation of the fluxes, although more PN information is required due to the $1/c$ scaling. Schematically, we use the expression of the $(\ell,m)$ mode~\eqref{eq:hlm} and use the relations of~\Cref{subsec:genmoments}, where we select the instantaneous contributions. Notice that contrary to the quasi-circular case, the multipole moments W and $\text{S}_i$ do contribute. Their expressions are given in Eqs.~(3.31) of~\cite{Dones:2024odv} for a generic motion. 
The LO (2,2) mode reads
\begin{align}
h_{22}^\text{inst} =& \frac{8 G \tmass \nu}{R\, c^4}\sqrt{\frac{\pi}{5}}e^{-2 \di \phi}\left\{\frac{G\tmass}{2r}-\frac{\rd^2}{2}+ \di r \rd \fid+\frac{r^2\fid^2}{2} \right. \nn\\
&\left. \qquad \qquad \qquad \qquad + \frac{9 G^5\tmass^5}{c^{10}r^5}\left[\frac{\mutp+ \delta\,\mutm}{\nu} \left( -\frac{G \tmass}{4r} -\rd^2 - \frac{4}{3} \di r \rd \fid + \frac{7}{12}r^2\fid^2   \right) + \frac{G \tmass}{r} \mutp \right]\right\}\,.
\end{align}
For all modes, we replace at consistent PN order $(r,\rd,\fid)$ by their expression in terms of $(x,\et,l)$ at $\calO(\et^{12})$, and we leave for now the phase $\phi$ untouched. We display here the LO to $\calO(\et^3)$
\begin{align}\label{eq:H22instxel}
H_{22}^\text{inst} =& 1 + \frac{\et}{4}\Bigl( e^{-\di l} + 5\, e^{\di l} \Bigr) +\frac{\et^2}{4}\Bigl( e^{-2\di l} - 2 + 7\, e^{2\di l}  \Bigr) + \frac{\et^3}{32}\Bigl( 9 e^{-3\di l} -5  e^{-\di l} - 33  e^{\di l} + 77  e^{3\di l} \Bigr) \nn \\
& + 3 \, x^5\Biggl\{ \frac{\mutp +\delta\,\mutm}{\nu} \left[ 1 + \frac{\et}{8}\Bigl( 47\, e^{-\di l} + 15 \, e^{\di l} \Bigr) + \frac{\et^2}{2}\Bigl( 38\, e^{-2\di l} + 19 + 6 \, e^{2\di l} \Bigr) \right. \nn \\ 
& \left. \qquad \qquad \qquad \qquad \qquad + \frac{\et^3}{64} \Bigl( 3073 e^{-3\di l} + 1655  e^{-\di l} + 919  e^{\di l} + 289  e^{3\di l}  \Bigr) \right]  \nn \\ 
& \qquad \qquad + \mutp \left[ 4 + \frac{\et}{2}\Bigl( 37\, e^{-\di l} + 35 \, e^{\di l} \Bigr) + \frac{\et^2}{4}\Bigl( 139\, e^{-2\di l} + 112 + 113 \, e^{2\di l} \Bigr) \right. \nn \\
& \qquad \qquad \qquad \qquad \left. + \frac{\et^3}{32} \Bigl( 1971 e^{-3\di l} + 3145  e^{-\di l} + 2833  e^{\di l} + 1163 e^{3\di l}  \Bigr) \right] \Biggr\} + \calO\left(\frac{1}{c^2},\frac{\etidal}{c^2}\right) + \calO\bigl(\et^4\bigr)\,.
\end{align}
The full expressions are available on demand. If it is found to be useful for waveform modeling purposes, it is also possible to express the modes in terms of $(x,\et,u)$ which has the good taste of being exact in eccentricity, although the expressions are more complex. To obtain it, one simply needs to replace the expressions of $(r,\rd,\fid)$ in terms of $(x,\et,u)$ derived in Paper~I, and insert them consistently in $h_{\ell m}(r,\rd,\phi,\fid)$. The expressions involving tides are very long, but the modes take the symbolic form at each PN order (point mass and tides)
\begin{equation}
H_{\ell m}^\text{inst}(x,\et,u) = \sum_k \frac{a_k}{(1-\et \cos u)^k} + \sum_{p} \frac{\di \, b_p\, \sin(u)}{(1-\et \cos u)^p}\,,
\end{equation}
where the coefficients $a$ and $b$ depend on the eccentricity, mass ratio and the tidal polarizabilities. Now, we turn to the tail contributions.

\subsection{Tail part}\label{subsec:modestail}

For the 2.5PN waveform, one needs to compute the tail contributions to $\text{U}_{ij}$ at NLO, and $\text{U}_{ijk}$, $\text{U}_{ijkl}$, $\text{V}_{ij}$ and $\text{V}_{ijk}$ at LO. We define the constant parameter $x_0'$ related to the gauge constant $b_0$ appearing in the tail integrals~\eqref{eq:tails} through the following relation
\begin{equation}
x_0'=\left( \frac{G\tmass}{c^3} \frac{e^{11/12-\gamma_E}}{4b_0}\right)^{2/3}.
\end{equation}
After writing the integrands of the time integrals as functions of $(x,\et,l,\lambda)$, we integrate them using~\eqref{eq:tailkernel}. Once again, the modes are obtained by projecting on the spherical harmonics basis. We obtain for the (2,2) mode at LO and $\calO(\et)$
\begin{align}\label{eq:h22tail}
H_{22}^\text{tail} =& \left\{ 2\pi + 6 \di \ln\left( \frac{x}{x_0'}\right) + \frac{\et}{4} \left[ e^{\di l}\left( 13\pi +39\di\ln\left( \frac{x}{x_0'}\right)+6\di\ln 2  \right)  + e^{-\di l} \left( 11\pi +33\di\ln\left( \frac{x}{x_0'}\right)+54\di\ln\left( \frac{3}{2}\right) \right)\right] \right\} x^{3/2}\,, \nn \\
& +\Biggl\{ \frac{\mutp + \delta \, \mutm}{\nu}\left[ 6\pi + 18 \di \ln\left( \frac{x}{x_0'}\right) + \frac{3\et}{8} \left[ e^{\di l} \left( 31\pi + 93 \di\ln\left( \frac{x}{x_0'}\right)+2\di\ln 2 \right) \right. \right. \nn \\
& \left. \left. + e^{-\di l} \left( 157\pi +471\di\ln\left( \frac{x}{x_0'}\right)+378\di\ln \left(\frac{3}{2}\right)  \right)\right] \right] \nn \\ 
& + \mutp \left[ 24\pi + 72 \di \ln\left( \frac{x}{x_0'}\right) + \frac{3\et}{4} \left[ e^{\di l} \left( 257\pi + 771 \di\ln\left( \frac{x}{x_0'}\right)+18\di\Bigl(23\ln 2 -5 \Bigr)\right) \right. \right. \nn \\
& \left. \left. + e^{-\di l} \left( 319\pi + 957 \di\ln\left( \frac{x}{x_0'}\right)+54\di\left(29 \ln \left(\frac{3}{2}\right)-5\right)  \right)\right] \right]\Biggr\} x^{13/2} + \calO \left( \frac{1}{c^5},\frac{\etidal}{c^5} \right) + \calO\bigl( \et^2 \bigr)\,.
\end{align}
The full results of this Section are available on demand. A good (although partial) check of these expressions together with the ones of~\Cref{sec:fluxes}, is to compute the fluxes from the modes using
\begin{equation}\label{eq:FGfromModes}
\mathcal{F} = \frac{c^3 R^2}{16 \pi G} \sum_{\ell=2}^\infty\sum_{m=-\ell}^\ell \vert \dot{h}_{\ell m}  \vert^2\,, \qquad \mathcal{G} = -\frac{c^3 R^2}{16 \pi G} \sum_{\ell=2}^\infty\sum_{m=-\ell}^\ell m \,\text{Im}\left[ \dot{h}_{\ell m} h^*_{\ell m}  \right]\,,
\end{equation}
where Im is the imaginary part and the star notation refers to the complex conjugate. Notice that we need to compute the time derivatives of the modes, which can be achieved in two ways: either redo a similar computation with the radiative moments derived one time, or start from the expressions~\eqref{eq:H22instxel}-\eqref{eq:h22tail} and apply the chain rule $\tfrac{\dd}{\dd t} = n\tfrac{\dd}{\dd l} + \Omega \tfrac{\dd }{\dd \lambda}$. Both methods have been used and yield the same results. Thus, we injected them in Eqs.~\eqref{eq:FGfromModes} and recovered the same expressions for the fluxes~\eqref{eq:FGtot} after orbit averaging and eccentricity expanding.

\subsection{Post-adiabatic corrections}\label{subsec:modesPA}

So far, we have used the conservative dynamics to derive the modes. However, one cannot neglect the PA contributions in the (2,2) and (2,0) modes because they are required at relative 2.5PN. We make use of the oscillatory PA corrections of $(\tilde{x},\tilde{e}_t,\tilde{l},\tilde{\lambda})$ derived in Section IV. of Paper I~\cite{paperI}. The PA corrections to the $(\ell,m)$ mode can be computed with
\begin{equation}\label{eq:hlmPA}
h_{\ell m}^\text{PA} = \frac{\partial h_{\ell m}}{\partial x} \tilde{x} + \frac{\partial h_{\ell m}}{\partial \et} \tilde{e}_t +\frac{\partial h_{\ell m}}{\partial l} \tilde{l} + \frac{\partial h_{\ell m}}{\partial \lambda} \tilde{\lambda} \,.
\end{equation}
Since they are relative 2.5PN quantities, we simply need to use the LO of the (2,2) and (2,0) modes expressed in terms of $(x,\et,l,\lambda)$ computed in Sec.~\ref{subsec:modesinst} to which we apply~\eqref{eq:hlmPA}. The PA contribution to the (2,2) mode up to $\calO\bigl(\et^3\bigr)$ reads
\begin{align}
H_{22}^\text{PA} =& \frac{192}{5}\di \nu x^{5/2} \Biggl\{ 1 + \frac{\et}{72}\biggl(401 e^{-\di l} + 293 e^{\di l} \biggr) +\frac{\et^2}{576}\biggl( 4391 e^{-2\di l} + 9248 + 4251 e^{2\di l} \biggr) \nn \\
& \qquad \qquad \qquad +\frac{\et^3}{27648}\biggl( 298895 e^{-3\di l} + 980196 e^{-\di l} + 878500 e^{\di l} + 330981 e^{3\di l} \biggr)\nn  \\
& + x^5\Biggl[ \frac{\mutp+\delta\,\mutm}{\nu}\left( 14+ \frac{\et}{768}\Bigl( 117351 e^{-\di l} + 53773 e^{\di l}\Bigr) + \frac{\et^2}{576}\Bigl( 281929 e^{-2\di l} + 331722 + 82509 e^{2\di l}\Bigr) \nn \right.\\
& \left. \qquad \qquad \qquad \qquad \qquad + \frac{\et^3}{18432}\Bigl(28437740 e^{-3\di l} +51685531 e^{-\di l} + 29533385 e^{\di l} + 5108656 e^{3\di l}\Bigr) \right) \\
& \qquad \quad + \mutp \left( 234+ \frac{\et}{64}\Bigl( 78963 e^{-\di l} + 57689 e^{\di l}\Bigr) + \frac{\et^2}{96}\Bigl( 227759 e^{-2\di l} + 508822 + 189469 e^{2\di l}\Bigr) \nn \right.\\
& \left. \qquad \qquad \qquad \qquad  + \frac{\et^3}{9216}\Bigl(39871745 e^{-3\di l} +181015338 e^{-\di l} + 137918590 e^{\di l} + 33535743e^{3\di l}\Bigr) \right)\Biggr]\Biggr\} +\calO\bigl(\et^4\bigr) \nn\,.
\end{align}
The quantities $(x,\et,l)$ must be understood in this expression as their secular part $(\bar{x},\bar{e}_t,\bar{l})$. We find agreement with the point-particle part given in~\cite{Boetzel:2019nfw} up to $\calO(\et^5)$, as explained in Paper~I, there is a discrepancy at $\calO(\et^6)$ due to an inconsistency in the derivation of $\tilde{l}$. For the tidal part, we recover the quasi-circular limit computed in~\cite{Dones:2024odv}.

\subsection{Waveform with phase redefinition}\label{subsec:modespsi}

Following what was done in~\cite{Boetzel:2019nfw}, we define new orbital elements
\begin{subequations}
\begin{align}
\xi &= \bar{l} - \frac{3 G \mathcal{M}}{c^3}\bar{n}\ln\left( \frac{\bar{x}}{x_0'}\right)\,,\\
\lambda_\xi &= \bar{\lambda} - \frac{3 G \mathcal{M}}{c^3}\bar{\Omega}\ln\left( \frac{\bar{x}}{x_0'}\right)\,,
\end{align}
\end{subequations}
where we recall that $\mathcal{M}$ is defined in~\eqref{eq:mADM}. These new variables are introduced to reabsorb the logarithms coming from the tail terms~\eqref{eq:h22tail}. The new phase variable corresponding to the new angles is simply the phase $\phi$ in which we replace $l$ by $\xi$ and $\lambda$ by $\lambda_\xi$, \textit{i.e.}\footnote{Note that in the presence of tidal effects, the structure of $W$ is such that Eq.~(71) of~\cite{Boetzel:2019nfw} is not valid here because the QKP including tidal effects is more complex.}
\begin{equation}
\psi = \lambda_\xi + W(\xi) = \bar{\lambda}_\xi + \tilde{\lambda}_\xi + \bar{W}(\xi) + \tilde{W}(\xi)\,.
\end{equation}
Since the modes are expressed in terms of the phase $\phi$, we can link the two phases by the relation, see~\cite{Boetzel:2019nfw} for more details. Finally, the amplitude modes can now be written in the following way
\begin{equation}
h_{\ell m} = \frac{8 G \tmass \nu \bar{x}}{R\, c^2}\sqrt{\frac{\pi}{5}} H^\psi_{\ell m} (\bar{x},\bar{e}_t,\xi) e^{-\di m \psi}\,,
\end{equation}
where the relative 1.5PN (2,2) mode at $\calO(\et)$ reads
\begin{align}
H^\psi_{22} =&\, 1 + \et\left( \frac{5}{4} e^{\di \xi} +\frac{1}{4}e^{-\di \xi} \right) + \left[-\frac{107}{42} +\frac{55}{42}\nu + \et \left( -\frac{31}{24}e^{\di \xi} -\frac{257}{168}e^{-\di\xi} +\nu\left( \frac{35}{24}e^{\di\xi}+\frac{169}{168}e^{-\di\xi}\right) \right)  \right]x \nn\\
& + \left[2\pi +\frac{\et}{4} \left( e^{\di \xi}\left(13\pi + 6\di \ln 2 \right) + e^{-\di \xi} \left(11\pi+54\di \ln (3/2)\right) \right) \right]x^{3/2}\nn\\
& + \left\{ \frac{\mutp+\delta\,\mutm}{\nu}\left[3 +\frac{\et}{8} \left(45 e^{\di\xi} + 141e^{-\di\xi} \right) \right] + \mutp\left[12 + \frac{\et}{2}\left( 105 e^{\di\xi} + 111e^{-\di\xi}\right)\right]\right\}x^5 \nn\\
& + \Biggl\{\frac{\mutp+\delta\,\mutm}{\nu}\left[\frac{9}{2} +\frac{125}{7}\nu +\et \left(33 e^{\di\xi}+\frac{111}{2}e^{-\di\xi} + \nu\left( \frac{12573}{112}e^{\di\xi} +\frac{11369}{112}e^{-\di\xi} \right)\right) \right]\nn\\
& +\mutp \left[-\frac{265}{7} +\frac{45}{7}\nu +\et \left(-\frac{263}{28} e^{\di\xi} - e^{-\di\xi}\left(\frac{3683}{28} -  \frac{153}{14} \nu\right)\right) \right] + \sigmatp\left[ \frac{224}{3} +\frac{\et}{3} \left( 1574e^{\di\xi} + 1262 e^{-\di\xi}\right) \right]\Biggr\}x^6 \nn\\
& + \left\{ \frac{\mutp+\delta\,\mutm}{\nu}\left[6\pi +\frac{3\et}{8} \left( e^{\di\xi}\Bigl( 31\pi +2\di \ln(2) \Bigr) + e^{-\di\xi}\Bigl( 157\pi +378\di \ln(3/2) \Bigr) \right) \right]  \right.\nn\\
& \left. + \mutp\left[24\pi +\frac{3\et}{4} \left( e^{\di\xi}\Bigl( 257\pi +18\di (23\ln(2) -5) \Bigr) + e^{-\di\xi}\Bigl( 319\pi +54\di ( 29\ln(3/2)-5) \Bigr) \right)\right]\right\}x^{13/2}\nn \\ & + \calO\bigl(\et^2\bigr) + \calO\left( \frac{1}{c^4},\frac{\etidal}{c^4} \right)\,.
\end{align}
A good consistency check is that all the logarithm contributions of $x_0'$ vanish in the final result, which is to be expected since it is a gauge constant. Comparing with the literature, we have checked that all the modes are in agreement with Eqs.~(4.13) of previous work~\cite{Dones:2024odv} in the limit $\et = 0$ at relative 2.5PN except for the (4,4) mode, which is the only one containing oscillatory memory terms in the quasi-circular limit, and the $m=0$ modes which contain the DC memory. We have checked that the difference at the level of the (4,4) mode uniquely comes from the memory. We find agreement for the point-particle part of the modes with~\cite{Ebersold:2019kdc}, except for the $\calO(\et^6)$ terms at 2.5PN. As explained in~\Cref{subsec:modesPA}, this is due to the fact that the derivation in this paper contains an inconsistency in the derivation of $\tilde{l}$ at $\calO(\et^6)$, because of a division by the eccentricity at leading order.  

Finally, we would like to recall that those results are formally valid up to $e_\text{max} \simeq 0.6627434$, although for a better accuracy, more eccentricity terms should be included if the value of the eccentricity is close to $e_\text{max}$. Indeed, in Section III.~C.~3. of Paper~I~\cite{paperI}, we discuss the radius of convergence of the power series in $\et$ of $u(l)$.

\section{Discussion on the phasing}\label{sec:num}

As we have seen in Paper~I, the perturbed dimensionless Binet equation in the presence of leading order tides is given by $y''+y=1+\varepsilon\, y^5$. The correction parameter $\varepsilon$ to the Keplerian motion reads
\begin{equation}\label{eq:epsilonval}
\varepsilon = \frac{6}{a^5(1-e^2)^5}\left( \frac{m_2}{m_1} k_1^{(2)}R_1^5 + \frac{m_1}{m_2} k_2^{(2)}R_2^5 \right) = \frac{\varepsilon_\text{circ}}{(1-e^2)^5}\,,
\end{equation}
where $a$, $e$ are the Newtonian semi-major axis and eccentricity, and $\varepsilon_\text{circ}$ is the perturbation in the case of circular orbits. One can expect that the perturbation of adiabatic tides on the waveform will be enhanced by the presence of eccentricity compared to the quasi-circular case. To illustrate it, let us turn to the phasing. As said in~\Cref{subsec:secularOrb}, the analytic resolution of the coupled system of differential equations~\eqref{eq:xdotedot} has been left for future works. In this discussion, we would like to get a grasp on the magnitude of the tidal eccentric terms compared to the eccentric point-particle case. To this end, we choose a very simple model where we consider a binary with an initial eccentricity $e_0$, an initial (GW) frequency of 20Hz, fixing the initial $x_0 = (G \tmass \pi f_\text{20Hz}/c^3)^{2/3}$. With these initial conditions, we evolve numerically the full 2.5PN system~\eqref{eq:xdotedot} beyond LO, neglecting the tidal current quadrupole and mass octupole contributions. This gives the secular evolution of $\bar{x}(t)$ and $\bar{e}_t(t)$. We consider four binaries with different parameters detailed in~\Cref{tab:cases}.  If one of the companions is a NS, we use the value of the compactness parameter $\mathcal{C}_A = G m_A/R_A c^2 = 0.15$. In particular, Case I corresponds to the inferred parameters of the NSBH that emitted the GW200105 signal~\cite{Morras:2025xfu}. The other three are hypothetical BNS. 
\begin{table}[htb]
	\begin{tabular}{|c||c|c|c|c|c|}
\hline  & $m_1 (M_\odot)$ & $m_2 (M_\odot)$ & $k_1^{(2)}$ & $k_2^{(2)}$ &  $e_0$\\ \hline 
\hline Case I & 11.5 & 1.5 & 0 & 0.1 & 0.14 \\ 
\hline Case II & 1.4 & 1.4 & 0.1 & 0.1  & 0.3 \\ 
\hline Case III & 1.4 & 1.4 & 0.1 & 0.1 & 0.6 \\
\hline Case IV & 1.8 & 0.8 & 0.1 & 0.1 & 0.3 \\\hline  
\end{tabular}
\caption{Case I is a NSBH with the inferred parameters of GW200105. Cases II and III are identical NSs of 1.4$M_\odot$, with different initial eccentricities. Case IV is a BNS of mass ratio $ 2.25$.}\label{tab:cases}
\end{table}
For each case, we stop the evolution at the reference frequency $x_\text{ref}=1/6$, which is chosen to match the Schwarzschild ISCO frequency in the case of quasi-circular orbits. We set the initial time $t_0=0$ and find the reference time by solving numerically the interpolated solution $\bar{x}(t_\text{ref})=1/6$. Next, we compute the phase (assuming $\phi(t_0)=0$), using $\phi = \lambda + W$ and considering only the secular part. One gets $ \dd\bar{\phi}/\dd t = \bar{\Omega}$, which yields
\begin{equation}
\bar{\phi}_\text{ref} = \frac{c^3}{G\tmass} \int_0^{t_\text{ref}}\dd t\, \bar{x}^{3/2}(t)\,.
\end{equation}
Finally, the number of GW cycles within the detector band is simply given by $N_\text{GW}=2 N_\text{orb}=\bar{\phi}_\text{ref}/\pi$. The different results for each case are given in~\Cref{tab:val}.
\begin{table}[htb]
	\begin{tabular}{|c||c|c|c|c|c|c|c|}
\hline $t_\text{ISCO} (s)$ & QC pp & QC pp \& tides & Ecc pp & Ecc pp \& tides & $\Delta\phi$ & $\Delta N$ \\ \hline 
\hline Case I & 30.1310 & 30.1309 & 28.1095 & 28.1095 & 0.041 & $0.013$ \\ 
\hline Case II & 160.737 & 160.734 & 113.054 & 113.051 & 7.54 & 2.40\\ 
\hline Case III & '' & '' & 31.2112 & 31.2072 & 7.60 & 2.42\\
\hline Case IV & 213.083 & 213.078 & 149.892 & 149.887 & 11.1 & 3.5\\ \hline  
\end{tabular}
\caption{For each case in~\Cref{tab:cases}, we solve the secular dynamics for $x(t)$ and $\et(t)$. In column one, we consider only $\dot{x}$ for zero eccentricity without tidal corrections. In column two, zero eccentricity with tidal corrections. In column three, with eccentricity without tidal corrections. In column four, with eccentricity with tidal corrections. Column five represents the difference of the accumulated phase $\Delta \phi$ within the detectors bands between the cases eccentric with and without tides, \textit{i.e.} between column three and four. Finally, column six represents the corresponding difference in GW cycles $\Delta N$.}\label{tab:val}
\end{table}
Let us comment first that the majority of the information is carried by the non-tidal eccentric PN secular dynamics. Next, we notice that for each case, the adiabatic tidal interaction shortens the time to reach the reference frequency by (at most) a few cycles. Then, focusing on Case~I, which corresponds to the NSBH case, we can see that $\Delta N$ is very small, probably due to the fact that $e_0$ is relatively small. However, in the case of BNSs, a high mass ratio or high initial eccentricity could lead to a non-negligible dephasing, which hints towards a possibly detectable effect in some regions of the parameter space.

Naturally, considering frequencies close to the ISCO is not realistic because we expect other matter effects to dominate around this frequency, such as dynamical tides, mass transfer, tidal disruption, electromagnetic fields... Specifically in the case of dynamical tides, the circularization in the late inspiral on eccentric orbits is expected to be mostly carried by the dissipation in the vibration modes of the NSs. Thus, the dephasing computed above constitute a lower limit to the ``full" dephasing considering dynamical tides on an eccentric motion.

To end the discussion, we would like to emphasize that this reasoning aims at illustrating the effects of adiabatic tides and a more thorough study is necessary to give any definite conclusions. For example, we have not considered the high bound of the frequency band of the detector. A convincing argument would be to consider two waveforms, one neglecting tides and the other including all information, and compute the mismatch. It would be interesting to look for next LIGO observing runs and Einstein Telescope.

\section{Summary and perspectives}\label{sec:ccl}

In Paper I~\cite{paperI}, we tackled the problem of the dynamics at the relative second-and-a-half PN order for a compact binary on eccentric orbits tidally interacting. We derived the conservative motion using a quasi-Keplerian parametrization and the radiation part of the dynamics. These results were used in the present paper to derive the radiated energy and angular momentum fluxes, as well as the GW amplitude modes to the same relative PN order. The memory parts of the angular momentum flux and the amplitude modes have been left for future works. We have also briefly discussed the effects of eccentric corrections to the tidal terms on the phasing. We found that in some particular cases, these new terms could affect the number of GW cycles in the detectors band, potentially leading to a non-negligible dephasing compared to the BBH case. The relevant results are gathered in the ancillary file~\cite{SuppMaterial2}, which contains: 
\begin{itemize}
\item the instantaneous fluxes in terms of $(r,\rd,\phi,\fid)$
\item the total orbit averaged fluxes: instantaneous and resummed tail at $\calO \bigl( \et^{14} \bigr)$
\item the secular evolution of the orbital elements $\langle \dot{x} \rangle$, $\langle \dot{e}_t \rangle$, $\langle \dot{k} \rangle$, $\langle \dot{n} \rangle$, $\langle \dot{a}_r \rangle$
\item the instantaneous part of the amplitude modes in terms of $(r,\rd,\phi,\fid)$
\item the modes $H_{\ell m}^\psi(x,\et,\xi)$ with the phase redefinition at $\calO \bigl( \et^{12} \bigr)$
\end{itemize} 

The next steps towards a better modeling of finite size effects in compact binaries are the following. The first one is to complete the derivation of the full waveform accounting for the oscillatory and DC memory effects. To this end, one would need to solve for $x(\et)$ analytically. This would also be a first step to solve analytically for the phasing. As, said in the discussion section, it would be interesting to perform a rigorous study to see whether eccentric corrections to adiabatic tides can be detectable for various current and future detectors. Even if it is not the case, this computation forms a first step towards the long term goal of having a clean description of dynamical tides if the system is on eccentric orbits which is the long term goal.

\section*{Acknowledgments}

Q.H. is grateful to Guillaume Faye, Anna Heffernan, François Larrouturou, Antoni Ramos-Buades and Lorenzo Speri for useful discussions. This work was supported by the Universitat de les Illes Balears (UIB); the Spanish Agencia Estatal de Investigación grants PID2022-138626NB-I00, RED2024-153978-E, RED2024-153735-E, funded by MICIU/AEI/10.13039/501100011033 and the ERDF/EU; and the Comunitat Autònoma de les Illes Balears through the Conselleria d'Educació i Universitats with funds from the European Union - NextGenerationEU/PRTR-C17.I1 (SINCO2022/6719) and from the European Union - European Regional Development Fund (ERDF) (SINCO2022/18146).

\appendix
\section{A computation of orbit average integrals}\label{app:Ikn}

The goal of this section is to compute the following integrals, $\forall (k,n)\in \mathbb{N}^2$,
\begin{subequations}\label{eqs:IknJkn}
\begin{align}
I_{k,n} &\equiv \int_0^{2\pi}\frac{\dd u}{2\pi} \frac{\cos{(ku)}}{(1-e\cos{u})^n} \,,\qquad\qquad J_{k,n} \equiv \int_0^{2\pi}\frac{\dd u}{2\pi} \frac{\cos{(ku)}}{(1-e\cos{u})^n}\ln(1-e \cos u)\,,\\
I'_{k,n} &\equiv \int_0^{2\pi}\frac{\dd u}{2\pi} \frac{\sin{(ku)}}{(1-e\cos{u})^n}  \,, \qquad\qquad  J'_{k,n} \equiv \int_0^{2\pi}\frac{\dd u}{2\pi} \frac{\sin{(ku)}}{(1-e\cos{u})^n}\ln(1-e \cos u)\,.
\end{align}
\end{subequations}
Remark that in this project, we do not encounter logarithms, however they appear at 3PN, which is why we deal with those here. Notice also that the integrals $I_{k,n}$ and $J_{k,n}$ can always be rewritten as combinations of $I_{0,n}$ and $J_{0,n}$ respectively. The expressions of $I_{0,n}$ and $J_{0,n}$ are well-known for $n\geq 1$, see \textit{e.g.} Eqs.~(545) of~\cite{Blanchet:2013haa}
\begin{subequations}\label{eq:IJ0nLuc}
\begin{align}
I_{0,n} &= \frac{1}{(1-e^2)^{n/2}}P_{n-1}\left( \frac{1}{\sqrt{1-e^2}}\right) = \frac{(-)^{n-1}}{(n-1)!}\left.\left[ \frac{\dd^{n-1}}{\dd z^{n-1}}\left(\frac{1}{\sqrt{z^2-e^2}}\right) \right]\right\rvert_{z=1}\label{eq:I0nLuc}\,,\\
J_{0,n} &= \frac{(-)^{n-1}}{(n-1)!}\left.\left[ \frac{\dd^{n-1}}{\dd z^{n-1}}\left(\frac{\mathcal{Z}(z,e)}{\sqrt{z^2-e^2}}\right) \right]\right\rvert_{z=1}, \quad \text{with} \quad \mathcal{Z}(z,e) = \ln\left[ \frac{1+\sqrt{1-\e^2}}{2} \right] + 2\ln\left[ 1+ \frac{\sqrt{1-e^2}-1}{z+\sqrt{z^2-e^2}} \right]\label{eq:J0nLuc}\,.
\end{align}
\end{subequations}
The problem with these expressions is that  if $n$ is sufficiently high, they become tedious to evaluate since one needs to take $n-1$ derivatives of non-trivial functions and evaluate them at $z=1$. In the case of tidal effects, the maximum $n$ required for orbit-averaging the fluxes is significantly higher than the point-particle case. Furthermore, these expressions do not cover the case $n=0$, while it was required in the computations of Paper~I for the post-adiabatic corrections of the orbital elements. Thus, we generalize these integrals and provide much more compact and convenient forms, in particular the new versions of Eqs.~\eqref{eq:IJ0nLuc} are given in~\eqref{eq:Iknval} and~\eqref{eq:J0n}. Firstly, let us focus on $I'_{k,n}$ and $J'_{k,n}$, one can trivially find that they vanish due to the odd behaviour of $\sin(k u)$. Next, in order to tackle the computation of $I_{k,n}$ and $J_{k,n}$ in the general case, we define the following functions in the complex plane
\begin{equation}\label{eq:Ikalphadef}
\forall(k,\alpha) \in \mathbb{N}\times\mathbb{C}, \quad I_k(\alpha) \equiv \int_0^{2\pi}\frac{\dd u}{2\pi} \frac{\cos{(ku)}}{(1-e\cos{u})^\alpha}\,.
\end{equation}
The elegant trick to compute $J_{k,n}$ is to remark that it can be deduced from the expression of $I_k$ using the relation $\tfrac{\partial}{\partial \alpha} (1-e\cos u)^{-\alpha} = - \tfrac{\ln(1-e\cos u)}{(1-e\cos u)^\alpha}$. This means that $\forall(k,n) \in \mathbb{N}^2$,
\begin{equation}\label{eq:IJknlim}
I_{k,n} = \lim_{\alpha\rightarrow n} I_k(\alpha)\,, \qquad \text{and} \qquad J_{k,n} = -\lim_{\alpha\rightarrow n} \frac{\partial I_k}{\partial \alpha}\,.
\end{equation}
So naturally, we will first compute $I_k$, then take its limit when $\alpha\rightarrow n\in\mathbb{N}$ to obtain $I_{k,n}$ and then deduce $J_{k,n}$ by computing its derivative with respect to $\alpha$ before taking the limit. Before diving in the computations, we recall the definition of the hypergeometric function ${}_p F_q$ which we will use throughout the proof
\begin{equation}
{}_p F_q(\alpha_1,\dots, \alpha_p;\beta_1,\dots,\beta_q;z) \equiv \sum_{k=0}^\infty \frac{(\alpha_1)_k\dots (\alpha_p)_k}{(\beta_1)_k\dots(\beta_q)_k}\frac{z^k}{k!}\,,
\end{equation}
where $(x)_k \equiv \tfrac{\Gamma(x+k)}{\Gamma(x)}$ are the Pochhammer symbols. We will extensively use the property of Pochhammer symbols of negative integers
\begin{equation}\label{eq:negPochh}
(-m)_i = \begin{cases}
    (-)^i\frac{m!}{(m-i)!} & \text{if $i \leq m$} \\
    0 & \text{otherwise}
  \end{cases}\,.
\end{equation}
which comes from the fact that $\forall j\in\mathbb{N}, \tfrac{1}{\Gamma(-j)} = 0$. We also would like to comment that the following procedure can also yield the values of $I_{k,n}$ and $J_{k,n}$ for negative values of $n$, but we do not consider this case here.

\subsubsection{Computation of $I_k(\alpha)$}

We pose $k\in \mathbb{N}$ and $\alpha\in \mathbb{C}$. We start from~\eqref{eq:Ikalphadef} and use the Chebyshev polynomials $\cos(ku) = T_k(\cos u)$ which we rewrite as a ${}_2F_1$ hypergeometric function. Then, we express $I_k(\alpha)$ by splitting the integral in two and posing $x = (1-\cos u)/2$, as 
\begin{equation}
I_k(\alpha) = \frac{(-)^k}{\pi (1+e)^\alpha}\int_0^1 \dd x\, x^{-1/2} (1-x)^{-1/2} \left[ 1-\left(\frac{2e}{1+e}\right)x\right]^{-\alpha} {}_2 F_1 \left(-k,k;\frac{1}{2};x\right)\,.
\end{equation}
This can be integrated using Eq. (7.512.9), p.813 of~\cite{gradshteyn2007}. We get
\begin{align}
I_k(\alpha)= \frac{(-)^k}{(1-e)^\alpha} \frac{1}{k!\, \Gamma(1-k)}\, {}_3F_2\left(1,\frac{1}{2},\alpha;1-k,k+1;-\frac{2e}{1-e}\right)\label{eq:intcosku}\,,
\end{align}
which is what was found in~\cite{Henry:2023tka} for $\alpha = n\in\mathbb{N}$. However, the current version of \textit{Mathematica} does not evaluate the specific values of the hypergeometric function ${}_3F_2$ for high values of $k$ or $n$. This expression can be expressed in a much simpler form. Indeed, one needs to remark that 
\begin{equation}
I_k(\alpha) = \frac{(-)^k}{(1-e)^\alpha k!} \lim_{a\rightarrow 1} \mathcal{F}_{k,\alpha}(a)\,, \qquad \text{with} \qquad \mathcal{F}_{k,\alpha}(a) = \frac{1}{\Gamma(a-k)}{}_3F_2\left(a, \frac{1}{2},\alpha;a-k,k+1;-\frac{2e}{1-e}\right)\,.
\end{equation}
Next, we use Eq. (7.4.1.2), p.497 of~\cite{PrudnikovVol3}\footnote{Notice that Eq.~(7.4.1.35),p.500 of~\cite{PrudnikovVol3}, which is the form of the ${}_3F_2$ functions that one can encounter in this computation, contains a typo: the upper bound of the sum should be $n-1$ instead of $n$.} to express the ${}_3F_2$ function as a sum of simpler ${}_2F_1$
\begin{equation}
\mathcal{F}_{k,\alpha}(a) = \sum_{\ell=0}^k (-z)^\ell \binom{k}{\ell} \frac{(1-a)_{k-\ell}}{\Gamma(a-k) (1-a)_k} \frac{(1/2)_\ell (\alpha)_\ell}{(k+1)_\ell} {}_2F_1\left(\ell+\frac{1}{2}, \alpha+\ell; k+\ell+1;z\right)\,,
\end{equation}
For $\ell \in \llbracket0,k\rrbracket$\footnote{This notation refers to the interval of integer numbers, explicitely $\llbracket a,b\rrbracket = \{ n \vert n \in\mathbb{Z} \cap [a,b] \}$.}, one can show that in the limit $a\rightarrow 1$, $\tfrac{(1-a)_{k-\ell}}{\Gamma(a-k) (1-a)_k} = (-)^k \delta_{k\ell}$ where $\delta_{k\ell}$ is the Kronecker symbol. This leads to
\begin{equation}
I_k(\alpha) = \frac{(\alpha)_k}{(1-e)^\alpha}\frac{(-z)^k}{2^{2k}\,k!} {}_2F_1\left(\alpha+k,k+\frac{1}{2};2k+1;z\right) \qquad\text{with}\qquad z= - \frac{2e}{1-e}\,.
\end{equation}
This expression can be further simplified by manipulating the hypergeometric functions, notably using Eq.~(7.3.1.68) p.457 and then Eq.~(7.3.1.4), p.454 of~\cite{PrudnikovVol3}. We finally get
\begin{equation}\label{eq:IkalphaHypergeom}
\forall(k,\alpha)\in\mathbb{N}\times\mathbb{C}, \quad I_k(\alpha) = \frac{(1-\sqrt{1-e^2})^k}{e^k\,k!}\frac{(\alpha)_k}{(1-e^2)^{\alpha/2}}\, {}_2F_1\left(\alpha,1-\alpha;k+1;-\frac{1-\sqrt{1-e^2}}{2\sqrt{1-e^2}}\right)\,.
\end{equation}
Furthermore, for $\mu\in\mathbb{C}\backslash \mathbb{N}^*$ and on the branch cut $z\in\mathbb{C}\backslash ]-\infty,1[$, the Legendre associated function of first kind can be defined in terms of hypergeometric functions as\footnote{See II.18., p.773 of~\cite{PrudnikovVol3}}
\begin{equation}\label{eq:Pmunuof2F1}
P^\mu_\nu(z) = \left( \frac{z+1}{z-1}\right)^{\mu/2}\frac{{}_2F_1\left(\nu+1,-\nu;1-\mu; (1-z)/2 \right)}{\Gamma(1-\mu)}\,.
\end{equation}
Thus, one can alternatively write $I_k$ in the following form
\begin{equation}\label{eq:IkalphaLegendre}
\forall(k,\alpha)\in\mathbb{N}\times\mathbb{C}, \quad I_k(\alpha) = \frac{(\alpha)_k}{(1-e^2)^{\alpha/2}}\, P_{\alpha-1}^{-k} \left(\frac{1}{\sqrt{1-e^2}}\right)\,.
\end{equation}
In the following, we will use both formulations~\eqref{eq:IkalphaHypergeom} and~\eqref{eq:IkalphaLegendre}. Note that since the argument of the Legendre associated function is superior to 1, one needs to implement this function using the type 3 in \textit{Mathematica}.

\subsubsection{Computation of $I_{k,n}$}

We are now able to take the limit $\alpha\rightarrow n\in \mathbb{N}$. In order to cover the whole set $(k,n)\in\mathbb{N}^2$, we need to split the cases $n=0$ and $n\geq 1$. For the case $n=0$, we use~\eqref{eq:IkalphaHypergeom} for $\alpha \rightarrow 0$, which leads to $\delta_{0k}$. For the case $n\geq 1$, we start from~\eqref{eq:IkalphaLegendre}, which gives
\begin{equation}\label{eq:IknLegendre}
I_{k,n} =\frac{1}{(1-e^2)^{n/2}} \frac{(n+k-1)!}{(n-1)!}P ^{-k}_{n-1}\left( \frac{1}{\sqrt{1-e^2}}\right)\,.
\end{equation}
This expression can also be recovered by directly integrating $I_{k,n}$ using the $b_k^n$ coefficients in Eqs.~(41) of~\cite{Boetzel:2017zza}. Indeed, one can show that for $n\geq 1$, $I_{0,n} = b_0^n$ and $I_{k\geq 0,n} = b_k^n/2$, this is due to the fact that the integrals $I_{k,n}$ actually correspond to the Fourier cosine coefficients of the function $(1-e\cos u)^{-n}$. Next, Eq.~\eqref{eq:IknLegendre} can be further simplified using the property of the Legendre associated function of a negative integer parameter which can be written as a finite sum (on the same branch cut as before)
\begin{equation}\label{eq:Psum}
P^{-k}_{n-1}(z) = \left( \frac{z-1}{z+1} \right)^{k/2} \,\sum_{\ell=0}^{n-1} \frac{\Gamma(n+\ell)}{\Gamma(n-\ell)}\frac{(z-1)^\ell}{2^\ell \ell!(k+\ell)!}\,,
\end{equation}
which is a consequence of~\eqref{eq:negPochh}. Finally, we find the result written in the very convenient form
\begin{subequations}
\begin{align}
\forall k \in \mathbb{N},\quad I_{k,0} &= \delta_{0k}\,,\\
\forall (k,n)\in \mathbb{N}\times\mathbb{N}^*, \quad I_{k,n}&= \frac{(n+k-1)!}{(n-1)!}\sum_{\ell=0}^{n-1}\frac{1}{2^\ell \ell! (k+\ell)!}\frac{(n+\ell-1)!}{ (n-\ell-1)!}\frac{(1-\sqrt{1-e^2})^{k+\ell}}{e^k(1-e^2)^{(n+\ell)/2}}
\,.\label{eq:Iknval}
\end{align}
\end{subequations}
This formula has been tested by comparing the integrated $I_{k,n}$ using \textit{Mathematica} for each specific values $k\in \llbracket0,15\rrbracket$ and $n\in \llbracket0,30\rrbracket$. Notice that the limit $e\rightarrow 0$ in~\eqref{eq:Iknval} is well defined since $\tfrac{1-\sqrt{1-e^2}}{e}=\tfrac{e}{1+\sqrt{1-e^2}}\rightarrow 0$, but it is simpler to think that $I_{k,n}(e=0) = \int \dd u \cos(ku)/2\pi = I_{k,0}$. We chose this formulation because it is the one that \textit{Mathematica} manages to simplify the most efficiently. The well-known result~\eqref{eq:I0nLuc} is simply the special case of~\eqref{eq:IknLegendre} for $k=0$.

\subsubsection{Computation of $J_{k,n}$}

In order to compute $J_{k,n}$, as we see in~\eqref{eq:IJknlim}, we need to take the derivative of $I_k$ with respect to $\alpha$ and take the limit $\alpha\rightarrow n$. Similarly to the previous integral, we treat separately the cases $n=0$ and $n\geq 1$. 
\begin{itemize}
    \item \textbf{Case} $n=0$ :
\end{itemize}
We use the form~\eqref{eq:IkalphaHypergeom} to compute the derivative of $I_k$
\begin{align}
\frac{\partial I_k}{\partial \alpha} &= \frac{(\alpha)_k \,E_k}{(1-e^2)^{\alpha/2}}\left[ {}_2F_1(\alpha,1-\alpha;k+1;-z') \left(  \psi(\alpha+k) -\psi(\alpha) - \ln\sqrt{1-e^2}\right) + \frac{\partial \,{}_2F_1(\alpha,1-\alpha;k+1;-z')}{\partial \alpha}\right]\,,
\end{align}
where $E_k = \tfrac{(1-\sqrt{1-e^2})^k}{e^k\,k!}$ and $z' = \tfrac{1-\sqrt{1-e^2}}{2\sqrt{1-e^2}}$. We used $\tfrac{\partial (\alpha)_k}{\partial \alpha} = (\alpha)_k(\psi(\alpha+k)-\psi(\alpha))$  with $\psi$ being the digamma function. Then, we need to compute the derivative of the hypergeometric function. We do so using the following relation $\tfrac{\partial (1-\alpha)_\ell}{\partial \alpha} = (1-\alpha)_\ell\bigl(\psi(1-\alpha)-\psi(1-\alpha+\ell)\bigr)$, which gives
\begin{equation}\label{eq:d2F1da}
\frac{\partial {}_2F_1(\alpha,1-\alpha;k+1;-z')}{\partial \alpha} = \sum_{\ell=1}^\infty \frac{(-z')^\ell}{\ell!(k+1)_\ell}\frac{\Gamma(\alpha+\ell)}{\Gamma(\alpha)}\frac{\Gamma(1-\alpha+\ell)}{\Gamma(1-\alpha)}\bigl( \psi(\alpha+\ell)-\psi(\alpha) -\psi(1-\alpha+\ell)+\psi(1-\alpha)\bigr)\,.
\end{equation}
The case $n=0$ amounts to take the limit $\alpha\rightarrow 0$ in~\eqref{eq:d2F1da}. We extensively use $1/\Gamma(\varepsilon) = \calO(\varepsilon)$ and seperating the cases $k=0$ and $k\geq 0$, we get
\begin{equation}\label{eq:Jk0}
J_{0,0} = \ln\left( \frac{1+\sqrt{1-e^2}}{2} \right)\,,\qquad \text{and} \qquad
\forall k\in\mathbb{N}^*, \quad J_{k,0} = - \frac{\bigl(1-\sqrt{1-e^2}\bigr)^k}{k\, e^k}\,.
\end{equation}
The case $k=0$ is consistent with the direct integration using Eq.(4.224.12) of~\cite{gradshteyn2007}. We also recover the integral~(A2b) of~\cite{Boetzel:2019nfw} which can be written $J_{0,0} - e J_{1,0}$. The formula~\eqref{eq:Jk0} has been tested with particular values computed with \textit{Mathematica} for $k\in\llbracket0,30\rrbracket$.

\begin{itemize}
    \item \textbf{Case} $n\geq 1$ :
\end{itemize}
Now we focus on the case $n\geq 1$. We recall that $z=\tfrac{1}{\sqrt{1-e^2}} \in ]1,\infty[ \subset \mathbb{C}\backslash [-1,1]$, we will use the formulas in that region of the complex plane. First, we start here from the form~\eqref{eq:IkalphaLegendre}, thus after taking the derivative and the limit $\alpha\rightarrow n$, we get
\begin{align}\label{eq:Jknnsup1}
J_{k,n} &= - \lim_{\alpha\rightarrow n} \frac{\partial I_k}{\partial \alpha} = -(n)_k z^n\left[ P^{-k}_{n-1} (z) \bigl( \ln\left(z\right) +\psi(n+k)-\psi(n) \bigr) +\left. \frac{\partial P^{-k}_{\nu}(z)}{\partial \nu} \right\rvert_{\nu=n-1} \right]\,,
\end{align}
The difficulty is to compute the derivative of the associated Legendre function with respect to its degree $\nu$. This problem has been widely tackled in~\cite{Szmytkowski2011}. To apply the formulas of that Ref, we need to separate the cases $k\leq n-1$ and $k\geq n$.

\begin{itemize}
\item[$\star$] \textbf{Subcase} $k\leq n-1$ :
\end{itemize}
We apply Eq.~(5.24) of~\cite{Szmytkowski2011} to~\eqref{eq:Jknnsup1}, which allows to rewrite the associated Legendre polynomials in terms of a positive $k$.
%
%
Then, we use~(5.7) of~\cite{Szmytkowski2011} that we have rewritten here in the more compact form (for $0\leq m\leq n$)
\begin{align}\label{eq:dpmunuinf}
\left. \frac{\partial P^{m}_{\nu}(z)}{\partial \nu} \right\rvert_{\nu=n} &= P^m_n(z) \left[ \ln\left(\frac{z+1}{2}\right) -\psi(n+1) - \psi(n-m+1)\right] \\
& +\frac{(n+m)!}{(n-m)!} \left(\frac{z-1}{z+1}\right)^{m/2} \sum_{k=0}^{n}\frac{(k+n)!\,\psi(k+n+1)}{2^k k!(k+m)!(n-k)!}(z-1)^k\left[ 1 + \frac{(k+m)!}{(k-m)!}\frac{(n-m)!}{(n+m)!}\left( \frac{z+1}{z-1}\right)^m\right]\,.\nn
\end{align}
We only need to replace the Legendre polynomial by its expression as a finite sum~\eqref{eq:Psum}, which in the end leads to the final result $\forall n\in\mathbb{N}^*, \forall k \in \llbracket0,n-1\rrbracket$,
\begin{align}\label{eq:Jknres}
J_{k,n} &= -\frac{(n+k-1)!}{(n-1)!}\sum_{\ell=0}^{n-1}\frac{1}{2^\ell \ell!(k+\ell)!}\frac{(n+\ell-1)!}{(n-\ell-1)!}\frac{\bigl(1-\sqrt{1-e^2}\bigr)^{\ell+k}}{e^k(1-e^2)^{(n+\ell)/2}}\nn\\
& \qquad \times \left[ \ln\left( \frac{1+\sqrt{1-e^2}}{2(1-e^2)}\right) + \left( 1+\frac{(\ell+k)!}{(\ell-k)!}\frac{(n-k-1)!}{(n+k-1)!}\frac{\bigl(1+\sqrt{1-e^2}\bigr)^{2k}}{e^{2k}}\right)\psi(n+\ell) - 2\psi(n) \right]\,.
\end{align}
This result can be written in many different forms, notably due to the fact that~\eqref{eq:dpmunuinf} can take different equivalent expressions as displayed in~\cite{Szmytkowski2011}. Although it is not obvious that the limit $e\rightarrow 0$ is well defined, we chose this formulation because it is the most compact. Note that this formula gives the non-recursive closed form of~\eqref{eq:J0nLuc} in the particular case $k=0$. To illustrate the simplification we display the formula, $\forall n\in\mathbb{N}^*$,
\begin{equation}\label{eq:J0n}
\int_0^{2\pi}\frac{\dd u}{2\pi} \frac{\ln(1-e\cos u)}{(1-e\cos u)^n} = -\sum_{\ell=0}^{n-1}\left[ \ln\left( \frac{1+\sqrt{1-e^2}}{2(1-e^2)}\right) + 2 \sum_{i=0}^{\ell-1}\frac{1}{n+i}\right]\frac{1}{2^\ell(\ell!)^2}\frac{(n+\ell-1)!}{(n-\ell-1)!}\frac{(1-\sqrt{1-e^2})^\ell}{(1-e^2)^{(n+\ell)/2}}\,.
\end{equation}
It has been tested with~\eqref{eq:J0nLuc} for $n\in \llbracket1,30\rrbracket$. Note that this new version of the result is much faster to evaluate.

\begin{itemize}
\item[$\star$] \textbf{Subcase} $k\geq n$ :
\end{itemize}
We have not found a nice compact form for this subcase. Since it is not required for orbit averaged integrals that we have to deal with, we do not display the result here. To derive it, one needs to apply any formula from Section 5.4.2 of~\cite{Szmytkowski2011} to Eq.~\eqref{eq:Jknnsup1}. A bit of work is required to combine the associated Legendre functions, and rewrite them as a finite sum using~\eqref{eq:Psum}. The final result involves the same functions as~\eqref{eq:Jknres} but in a more complex form.

\section{Lengthy results}\label{app:lengthy}
\subsection{Orbit averaged fluxes}\label{app:fluxOrbAvg}

We display here the PN coefficients of Eqs.~\eqref{eq:FGtot}. We recall that the even PN terms are the instantaneous contributions which are exact in eccentricity, while the odd terms are the tail contributions which are resumed and valid to $\calO(\et^{14})$
\begin{subequations}
\begin{align}
\mathcal{F}_{0} =&\, 1+\frac{73}{24} \et^2+\frac{37}{96}\et^4\,,\\
\mathcal{F}_{1} =&\, -\frac{1247}{336} +\frac{10475}{672}\et^2 +\frac{10043}{384}\et^4 +\frac{2179}{1792}\et^6 +\nu\left( -\frac{35}{12} -\frac{1081}{36}\et^2 -\frac{311}{12}\et^4 -\frac{851}{576}\et^6 \right)\,,\\
\mathcal{F}_{1.5} =&\, 4+\frac{1375}{48}\et^2 + \frac{3935}{192}\et^4 + \frac{10007}{9216}\et^6 + \frac{2321}{221184}\et^8 - \frac{237857}{88473600}\et^{10} + \frac{182863}{1061683200}\et^{12} + \frac{4987211}{1664719257600}\et^{14} 
\,,\\
\mathcal{F}_{2} =&\, -\frac{203471}{9072} -\frac{3807197}{18144}\et^2 -\frac{268447}{24192}\et^4 +\frac{1307105}{16128}\et^6 +\frac{86567}{64512}\et^8 +\nu\left( \frac{12799}{504} +\frac{116789}{2016}\et^2 -\frac{2465027}{8064}\et^4 \right. \nn\\
& \left. -\frac{416945}{2688}\et^6 -\frac{9769}{4608}\et^8 \right) +\nu^2\left(\frac{65}{18} +\frac{5935}{54}\et^2 +\frac{247805}{864}\et^4 +\frac{185305}{1728}\et^6 +\frac{21275}{6912}\et^8  \right)\nn\\
& + \sqrt{1-\et^2}\left[ \frac{35}{2} +\frac{6425}{48}\et^2 + \frac{5065}{64}\et^4 +\frac{185}{96}\et^6 + \nu \left( -7 -\frac{1285}{24}\et^2 - \frac{1013}{32}\et^4 -\frac{37}{48}\et^6\right)\right]\,,\\
\mathcal{F}_{2.5} =& -\frac{8191}{672} -\frac{583}{24}\nu + \et^2 \left( \frac{36067}{336} - \frac{717733}{2016}\nu \right) + \et^4 \left( \frac{19817891}{43008} - \frac{21216061}{32256}\nu \right) + \et^6 \left( \frac{62900483}{387072} - \frac{78753305}{387072}\nu \right) \nn\\
& + \et^8 \left( \frac{26368199}{7077888} - \frac{208563695}{37158912}\nu \right) - \et^{10} \left( \frac{1052581}{34406400} - \frac{46886227}{3715891200}\nu \right) + \et^{12} \left( \frac{686351417}{95126814720} - \frac{151928969}{50960793600}\nu \right) \nn\\
& - \et^{14} \left( \frac{106760742311}{69918208819200} - \frac{1053619211}{2796728352768}\nu \right)   \,,\\
\mathcal{F}_{5} =& \Bigl(\mutp+ \delta\,\mutm \Bigr)\left[ 1 + \frac{211}{8}\et^2+\frac{3369}{32}\et^4+\frac{6275}{64}\et^6+\frac{10355}{512}\et^8 + \frac{225}{512}\et^{10}\right] + \nu \, \mutp \left[-3 + \frac{1247}{12}\et^2+\frac{56069}{192}\et^4 \right. \nn\\
& \left. + \frac{5341}{32}\et^6  + \frac{42019}{3072}\et^8 + \sqrt{1-\et^2}\left( 7 + \frac{1327}{24}\et^2+\frac{1081}{24}\et^4+\frac{3335}{384}\et^6+\frac{37}{192}\et^8\right) \right] \nn\,,\\
\mathcal{F}_{6} =&\Bigl(\mutp+ \delta\,\mutm \Bigr)\left\{ -\frac{22}{21} + \frac{65333}{672}\et^2+\frac{3487097}{2688}\et^4+\frac{8033143}{2688}\et^6+\frac{35957375}{21504}\et^8 + \frac{17296331}{86016}\et^{10}+\frac{255175}{114688}\et^{12} \nn \right.\\
&  \left. + \nu \left[ -\frac{373}{24} - \frac{1687}{24}\et^2 -\frac{276313}{256}\et^4 -\frac{2104931}{768}\et^6 -\frac{20956285}{12288}\et^8 -\frac{2930315}{12288}\et^{10}-\frac{3675}{1024}\et^{12} \right. \right. \nn \\
& \left. \left. + \sqrt{1-\et^2}\left( \frac{175}{12} + \frac{35275}{288}\et^2+\frac{9575}{64}\et^4+\frac{235325}{4608}\et^6+\frac{925}{768}\et^8 \right)\right]\right\} \nn \\
&  +\nu \, \mutp \left\{ -\frac{5195}{112} - \frac{85059}{448}\et^2+\frac{6358111}{1344}\et^4+\frac{47757523}{5376}\et^6+\frac{141282433}{43008}\et^8 + \frac{30408793}{172032} \et^{10} + \frac{5225}{57344} \et^{12} \right. \nn \\
& \left. + \nu \left( \frac{355}{12} - \frac{4121}{6}\et^2- \frac{521233}{96}\et^4-\frac{4385885}{576}\et^6 -\frac{6150043}{2304}\et^8 - \frac{473983}{3072} \et^{10} \right) \right.  \nn \\
&\left. + \sqrt{1-\et^2}\left[ \frac{4491}{112} + \frac{413885}{448}\et^2+\frac{1802267}{672}\et^4+\frac{15524179}{10752}\et^6+\frac{3330293}{21504}\et^8 + \frac{1591}{768} \et^{10} \right. \right. \nn \\
&  \left. \left. + \nu \left( -\frac{665}{12} - \frac{59675}{72}\et^2- \frac{366875}{192}\et^4-\frac{549985}{576}\et^6 -\frac{177415}{1536}\et^8 - \frac{3145}{2304} \et^{10} \right)\right] \right\} \nn\,,\\
&+ \Bigl(\sigmatp+ \delta\,\sigmatm \Bigr)\left[ -\frac{1}{9} - \frac{29}{3}\et^2-\frac{761}{8}\et^4-\frac{889}{4}\et^6-\frac{18445}{128}\et^8 - \frac{2905}{128}\et^{10} -\frac{1225}{3072}\et^{12}\right] \nn \\
&  +\nu \, \sigmatp \left[-\frac{344}{3} + \frac{7759}{18}\et^2+\frac{240697}{72}\et^4+\frac{23291}{6}\et^6+\frac{1141961}{1152}\et^8 +\frac{31261}{2304}\et^{10}-\frac{175}{192}\et^{12} \right. \nn\\
& \left. + \sqrt{1-\et^2}\left( 140 + \frac{7685}{6}\et^2+\frac{18055}{8}\et^4+\frac{26375}{24}\et^6+\frac{6545}{64}\et^8+\frac{185}{96}\et^{10}\right) \right] \,,\\
\mathcal{F}_{6.5} =& \Bigl(\mutp+ \delta\,\mutm \Bigr)\biggl[4 + \frac{5735}{32}\et^2 + \frac{5115}{4}\et^4 + \frac{44554019}{18432}\et^6 + \frac{98557247}{73728}\et^8 + \frac{10920042667}{58982400}\et^{10} + \frac{172228891}{58982400}\et^{12}\nn \\
&  + \frac{504548129}{1109812838400}\et^{14} \biggr]  + \nu\mutp\biggl[16 + \frac{102895}{96}\et^2 + \frac{377305}{96}\et^4 + \frac{22863647}{6144}\et^6 + \frac{3041353}{3456}\et^8 - \frac{3344063513}{35389440}\et^{10} \nn \\
&  - \frac{9177435133}{106168320}\et^{12} - \frac{7725639632419}{133177540608}\et^{14} \biggr]
\,,\\
\mathcal{F}_{7} =& \Big(\mutp+\delta\mutm\Big) \bigg\{- \frac{1064}{27} - \frac{177298}{189} \et^{2} + \frac{1251149}{1344} \et^{4} + \frac{784069}{32} \et^{6} + \frac{1187208755}{32256} \et^{8} + \frac{140301263}{10752} \et^{10} \nn\\
& + \frac{488121623}{516096} \et^{12} + \frac{199025}{43008} \et^{14} + \left[ - \frac{557447}{4032} + \frac{7313857}{16128} \et^{2} + \frac{152481845}{21504} \et^{4} - \frac{1185108787}{64512} \et^{6} - \frac{832131471}{16384} \et^{8} \right. \nn\\
& \left.  - \frac{5135754241}{229376} \et^{10} - \frac{970016633}{516096} \et^{12} - \frac{8377025}{688128} \et^{14}\right]\nu + \left[\frac{6853}{72} + \frac{5701}{288} \et^{2} + \frac{2008405}{2304} \et^{4} + \frac{45799997}{2304} \et^{6}\right. \nn\\
& \left. + \frac{1260800339}{36864} \et^{8} + \frac{14936425}{1024} \et^{10} + \frac{11735545}{8192} \et^{12} + \frac{62475}{4096} \et^{14}\right]\nu^{2} + \sqrt{1-\et^{2}}\left[\frac{85}{2} + \frac{25625}{16} \et^{2} + \frac{606665}{64} \et^{4} \right. \nn\\
& \left. + \frac{1839305}{128} \et^{6} + \frac{6138775}{1024} \et^{8} + \frac{570625}{1024} \et^{10} + \frac{1125}{256} \et^{12} + \left(\frac{78557}{448} + \frac{1976855}{768} \et^{2} + \frac{34619653}{5376} \et^{4} + \frac{40437367}{21504} \et^{6} \right. \right. \nn \\\
& \left.\left. - \frac{74750315}{86016} \et^{8} - \frac{4200695}{21504} \et^{10} - \frac{225}{128} \et^{12}\right)\nu  + \left(- \frac{5495}{48} - \frac{177125}{96} \et^{2} - \frac{4100765}{768} \et^{4} - \frac{18600415}{4608} \et^{6} - \frac{15578975}{18432} \et^{8} \right. \right. \nn \\\
& \left.\left. - \frac{37925}{3072} \et^{10}\right)\nu^{2}\right] \biggr\} + \nu\,\mutp  \bigg\{ \frac{2398157}{6048} - \frac{281963089}{36288} \et^{2} - \frac{5548399357}{145152} \et^{4} + \frac{4212096883}{145152} \et^{6} + \frac{19778967829}{258048} \et^{8} \nn \\\
&  + \frac{508139729}{24576} \et^{10} + \frac{442037275}{688128} \et^{12} + \frac{260375}{688128} \et^{14} + \left[\frac{156607}{672} + \frac{1282973}{192} \et^{2} - \frac{14354111}{576} \et^{4} - \frac{3074724907}{16128} \et^{6}\right. \nn \\\
& \left. - \frac{3100576597}{16128} \et^{8} - \frac{23602524377}{516096} \et^{10} - \frac{1175794961}{688128} \et^{12} - \frac{245575}{344064} \et^{14}\right]\nu + \left[- \frac{2675}{24} + \frac{18665}{12} \et^{2} + \frac{20989805}{576} \et^{4} \right. \nn \\\
& \left. + \frac{42545933}{384} \et^{6} + \frac{282370283}{3072} \et^{8} + \frac{131423089}{6144} \et^{10} + \frac{65991449}{73728} \et^{12}\right]\nu^{2} + \sqrt{1-\et^{2}}\left[- \frac{8517371}{18144} + \frac{32902283}{4536} \et^{2} \right. \nn \\\
& \left. + \frac{4499374025}{72576} \et^{4} + \frac{25011380305}{290304} \et^{6} + \frac{5332239499}{193536} \et^{8} + \frac{101567561}{64512} \et^{10} + \frac{1128661}{129024} \et^{12} + \left(- \frac{47911}{288} - \frac{6220115}{504} \et^{2} \right. \right. \nn \\\
& \left. \left. - \frac{568952171}{8064} \et^{4} - \frac{210413459}{2304} \et^{6} - \frac{1248269213}{43008} \et^{8} - \frac{162801595}{86016} \et^{10} - \frac{32609}{3072} \et^{12}\right)\nu + \left(\frac{1535}{8} + \frac{85025}{16} \et^{2} \right. \right. \nn \\\
& \left. \left. + \frac{4756115}{192} \et^{4} + \frac{3792785}{128} \et^{6} + \frac{7341545}{768} \et^{8} + \frac{383745}{512} \et^{10} + \frac{5365}{1536} \et^{12}\right)\nu^{2}\right] \bigg\} + \Big(\sigmatp+\delta\sigmatm\Big) \bigg\{- \frac{173}{756} - \frac{29867}{1512} \et^{2} \nn\\
& - \frac{121617}{224} \et^{4} - \frac{11351689}{4032} \et^{6} - \frac{17371873}{4608} \et^{8} - \frac{4144513}{3072} \et^{10} - \frac{1179323}{12288} \et^{12} - \frac{4975}{24576} \et^{14} + \sqrt{1-\et^{2}}\left(\frac{833}{3} \right. \nn\\
& \left. + \frac{202895}{72} \et^{2} + \frac{1940771}{288} \et^{4} + \frac{107933}{24} \et^{6} + \frac{1984325}{2304} \et^{8} + \frac{22015}{1152} \et^{10}\right)\nu + \left(- \frac{6475}{27} + \frac{49571}{72} \et^{2} + \frac{8825623}{864} \et^{4} \right. \nn\\
& \left.  + \frac{1946323}{96} \et^{6} + \frac{187210337}{13824} \et^{8} + \frac{90737401}{27648} \et^{10} + \frac{5009293}{18432} \et^{12} + \frac{57575}{18432} \et^{14}\right)\nu\bigg\} + \nu\sigmatp \biggl\{ - \frac{234977}{756} - \frac{1449269}{216} \et^{2} \nn\\
& + \frac{23328337}{672} \et^{4} + \frac{93868111}{672} \et^{6} + \frac{3383651893}{32256} \et^{8} + \frac{374265211}{21504} \et^{10} + \frac{55457981}{258048} \et^{12} - \frac{26875}{6144} \et^{14} + \left[\frac{2939}{3} \right. \nn \\
& \left. + \frac{19553}{18} \et^{2} - \frac{3431815}{72} \et^{4} - \frac{9166039}{72} \et^{6} - \frac{98713093}{1152} \et^{8} - \frac{33237985}{2304} \et^{10} - \frac{1201483}{9216} \et^{12} + \frac{8225}{1152} \et^{14}\right]\nu \nn \\
& + \sqrt{1-\et^{2}}\left[\frac{26989}{84} + \frac{2059525}{126} \et^{2} + \frac{923879}{12} \et^{4} + \frac{335595709}{4032} \et^{6} + \frac{411339925}{16128} \et^{8} + \frac{15467135}{10752} \et^{10} + \frac{7955}{576} \et^{12} \right. \nn \\
& \left. + \left(- \frac{3563}{3} - \frac{176620}{9} \et^{2} - \frac{401677}{6} \et^{4} - \frac{9657623}{144} \et^{6} - \frac{11465375}{576} \et^{8} - \frac{478465}{384} \et^{10} - \frac{3515}{288} \et^{12}\right)\nu\right]\bigg\}\nn \\
& + \nu\,\muthp \bigg[ - \frac{365}{6} + \frac{120055}{144} \et^{2} + \frac{2820745}{576} \et^{4} + \frac{95165}{16} \et^{6} + \frac{16942505}{9216} \et^{8} + \frac{1848025}{18432} \et^{10} \nn \\ 
& + \sqrt{1-\et^{2}}\left(\frac{175}{2} + \frac{38425}{48} \et^{2} + \frac{90275}{64} \et^{4} + \frac{131875}{192} \et^{6} + \frac{32725}{512} \et^{8} + \frac{925}{768} \et^{10}\right)\bigg] \,,\\
\mathcal{F}_{7.5} =& \Bigl(\mutp+\delta\,\mutm\Bigr)\bigg[ - \frac{88}{21} + \frac{582779}{1344}\et^{2} + \frac{60118129}{5376}\et^{4} + \frac{38016676877}{774144}\et^{6} + \frac{369661480627}{6193152}\et^{8} + \frac{52635331444483}{2477260800}\et^{10}\nn\\
& + \frac{10757510608133}{5945425920}\et^{12} + \frac{738699024014411}{46612139212800}\et^{14} -\nu\left( \frac{351}{16} + \frac{444379}{1152}\et^{2} + \frac{1055962981}{64512}\et^{4} + \frac{101194095859}{1548288}\et^{6} \right. \nn\\
& \left. + \frac{5579159415841}{74317824}\et^{8} + \frac{402060263746273}{14863564800}\et^{10} + \frac{1977828625007563}{713451110400}\et^{12} + \frac{6596529084276203}{31074759475200}\et^{14} \right) \bigg] \nn\\
&  +\nu\,\mutp \bigg[  \frac{1159}{168} + \frac{706813}{112}\et^{2} + \frac{395455463}{5376}\et^{4} + \frac{65074287517}{387072}\et^{6} + \frac{1287546737921}{12386304}\et^{8} + \frac{3917405226499}{309657600}\et^{10}\nn\\
& - \frac{612306686609083}{118908518400}\et^{12} - \frac{258024606036798199}{69918208819200}\et^{14} -\nu\left( \frac{2029}{12} + \frac{57816431}{4032}\et^{2} + \frac{1510120273}{16128}\et^{4} \right. \nn\\
&\left. + \frac{14106910105}{86016}\et^{6} + \frac{1713238113619}{18579456}\et^{8} + \frac{78867082008107}{7431782400}\et^{10} - \frac{780577617049577}{178362777600}\et^{12} \right. \nn\\
& \left. - \frac{17440100047953143}{5593456705536}\et^{14} \right) \bigg] - \Bigl(\sigmatp+\delta\,\sigmatm\Bigr)\bigg[  \frac{2}{9} + \frac{121}{3}\et^{2} + \frac{102791}{144}\et^{4} + \frac{7957087}{2592}\et^{6} + \frac{454728761}{110592}\et^{8} \nn\\
& + \frac{29124113557}{16588800}\et^{10} + \frac{31516563373}{159252480}\et^{12} + \frac{7757659367}{2890137600}\et^{14} \bigg] +\nu\,\sigmatp \bigg[ \frac{904}{9} + \frac{97399}{9}\et^{2} + \frac{403186}{9}\et^{4} \nn\\
& + \frac{265707241}{5184}\et^{6} + \frac{118703167}{82944}\et^{8} - \frac{195662101247}{16588800}\et^{10} - \frac{275797798033}{49766400}\et^{12} - \frac{922169068508227}{312134860800}\et^{14} \bigg] \,.
\end{align}
\end{subequations}
and
\begin{subequations}
\begin{align}
\mathcal{G}_{0} =&\, 1+\frac{7}{8} \et^2\,,\\
\mathcal{G}_{1} =&\, -\frac{1247}{336} + \frac{3019}{336}\et^2 + \frac{8399}{2688}\et^4 + \nu\left( -\frac{35}{12} -\frac{335}{24}\et^2-\frac{275}{96}\et^4 \right) \,,\\
\mathcal{G}_{1.5} =&\, 4+\frac{97}{8}\et^2 + \frac{49}{32}\et^4 - \frac{49}{4608}\et^6 - \frac{109}{36864}\et^8 - \frac{2567}{14745600}\et^{10} + \frac{4649}{176947200}\et^{12} + \frac{418837}{55490641920}\et^{14} \,,\\
\mathcal{G}_{2} =&\, -\frac{135431}{9072} - \frac{598435}{6048}\et^2 + \frac{30271}{3456}\et^4 + \frac{30505}{16128}\et^6+\nu \left( \frac{11287}{504} + \frac{9497}{672}\et^2 -\frac{106381}{1344}\et^4 - \frac{2201}{448}\et^6\right)\nn \\
& +\nu^2 \left( \frac{65}{18} + \frac{773}{12}\et^2 + \frac{569}{8}\et^4 + \frac{1519}{288}\et^6\right) + \sqrt{1-\et^2}\left[ 10 + \frac{335}{8}\et^2 + \frac{35}{8} \et^4 + \nu\left( -4 -\frac{67}{4}\et^2 - \frac{7}{4}\et^4 \right)\right]\,,\\
\mathcal{G}_{2.5} =& -\frac{8191}{672} -\frac{583}{24}\nu + \et^2 \left( \frac{108551}{1344} - \frac{32821}{168}\nu \right) + \et^4 \left( \frac{5055125}{43008} - \frac{1566125}{10752}\nu \right) + \et^6 \left( \frac{4125385}{774144} - \frac{712219}{96768}\nu \right) \nn\\
& - \et^8 \left( \frac{11065099}{49545216} - \frac{457507}{12386304}\nu \right) + \et^{10} \left( \frac{68397463}{2477260800} - \frac{792569}{309657600}\nu \right) + \et^{12} \left( \frac{194038163}{95126814720} - \frac{24946181}{39636172800}\nu \right) \nn\\
& - \et^{14} \left( \frac{3310841491}{15537379737600} - \frac{30761471}{647390822400}\nu \right)  \,,\\
\mathcal{G}_{5} =& \Bigl( \mutp+\delta\,\mutm\Bigr)\left[ 1 + \frac{117}{8}\et^2+\frac{915}{32}\et^4+\frac{635}{64}\et^6+\frac{165}{512}\et^8 \right] \nn \\
& + \nu \,\mutp \left[ \frac{251}{4}\et^2+\frac{5625}{64}\et^4+\frac{143}{8}\et^6+\frac{15}{64}\et^8 + \sqrt{1-\et^2}\left( 4 + \frac{71}{4}\et^2+\frac{95}{16}\et^4+\frac{7}{16}\et^6\right)\right]\,,\\
\mathcal{G}_{6} =&\Bigl(\mutp+ \delta\,\mutm \Bigr)\left\{ -\frac{22}{21} + \frac{17061}{224}\et^2+\frac{57741}{112}\et^4+\frac{483485}{896}\et^6+\frac{45195}{448}\et^8 + \frac{44175}{28672}\et^{10} \nn \right.\\
& \qquad \qquad \left. + \nu \left[ -\frac{223}{24} - \frac{2497}{48}\et^2 -\frac{109909}{256}\et^4 -\frac{424765}{768}\et^6 -\frac{772225}{6144}\et^8 -\frac{2815}{1024}\et^{10}\right. \right. \nn \\
& \qquad \qquad\qquad \left. \left. + \sqrt{1-\et^2}\left( \frac{25}{3} + \frac{1975}{48}\et^2+\frac{5725}{192}\et^4+\frac{175}{64}\et^6\right)\right]\right\}  \\
&  +\nu \, \mutp \left\{ -\frac{1457}{56} + \frac{14867}{224}\et^2+\frac{432973}{192}\et^4+\frac{1735103}{896}\et^6 +\frac{22218397}{86016}\et^8 + \frac{32955}{14336} \et^{10} \right. \nn \\
&  \qquad \qquad \left. + \nu \left( \frac{25}{3} - \frac{4455}{8}\et^2- \frac{145405}{64}\et^4-\frac{154211}{96}\et^6 -\frac{102903}{512}\et^8 - \frac{215}{128} \et^{10} \right) \right.  \nn \\
&\qquad \qquad  \left. + \sqrt{1-\et^2}\left[ \frac{1105}{56} + \frac{11187}{32}\et^2+\frac{287101}{448}\et^4+\frac{220961}{1792}\et^6+\frac{301}{64}\et^8 \right. \right. \nn \\
&  \qquad \qquad \left. \left. + \nu \left( -\frac{205}{6} - \frac{995}{3}\et^2- \frac{3445}{8}\et^4-\frac{8405}{96}\et^6 -\frac{595}{192}\et^8 \right)\right] \right\} \nn\,,\\
&+ \Bigl(\sigmatp+ \delta\,\sigmatm \Bigr)\left[ -\frac{1}{9} - \frac{9}{2}\et^2-\frac{91}{4}\et^4-\frac{595}{24}\et^6-\frac{735}{128}\et^8 - \frac{35}{256}\et^{10} \right]  \\
&  +\nu \, \sigmatp \left[-\frac{164}{3} + \frac{1342}{3}\et^2+\frac{3373}{2}\et^4+\frac{2926}{3}\et^6+\frac{46957}{384}\et^8 +\frac{5}{4}\et^{10} \right. \nn\\
& \left. \qquad \qquad + \sqrt{1-\et^2}\left( 80 +455\et^2 +\frac{1095}{2}\et^4+\frac{755}{8}\et^6+\frac{35}{8}\et^8\right) \right] \nn\,,\\
\mathcal{G}_{6.5} =& \Bigl(\mutp+ \delta\,\mutm \Bigr)\biggl[4 + \frac{1689}{16}\et^2 + \frac{13509}{32}\et^4 + \frac{3633127}{9216}\et^6 + \frac{1002787}{12288}\et^8 + \frac{52964281}{29491200}\et^{10} - \frac{364057}{58982400}\et^{12}\nn \\
&  + \frac{1644019}{822083584}\et^{14} \biggr]  + \nu\mutp\biggl[16 + \frac{4993}{8}\et^2 + \frac{79397}{64}\et^4 + \frac{1352599}{2304}\et^6 + \frac{21139}{73728}\et^8 - \frac{75855053}{2949120}\et^{10} \nn \\
&  - \frac{1119428567}{70778880}\et^{12} - \frac{39047851319}{3468165120}\et^{14} \biggr]
\,,\\
\mathcal{G}_{7} =&  \Big(\mutp+\delta\mutm\Big) \bigg\{- \frac{1723}{54} - \frac{342487}{1008} \et^{2} + \frac{15937165}{8064} \et^{4} + \frac{1706433}{224} \et^{6} + \frac{72681235}{16128} \et^{8} + \frac{60784667}{129024} \et^{10} + \frac{984905}{344064} \et^{12} \nn\\
&  + \left[ - \frac{26305}{504} + \frac{3960659}{8064} \et^{2} + \frac{48317519}{21504} \et^{4} - \frac{10224587}{1536} \et^{6} - \frac{2432242069}{344064} \et^{8} - \frac{41475991}{43008} \et^{10} - \frac{1000485}{114688} \et^{12} \right]\nu  \nn\\
& + \left[\frac{7391}{144} - \frac{19549}{144} \et^{2} + \frac{369833}{576} \et^{4} + \frac{14223331}{2304} \et^{6} + \frac{96611165}{18432} \et^{8} + \frac{7419055}{9216} \et^{10} + \frac{49575}{4096} \et^{12} \right]\nu^{2} \nn\\
&  + \sqrt{1-\et^{2}}\left[35 + \frac{3205}{4} \et^{2} + \frac{85215}{32} \et^{4} + \frac{3875}{2} \et^{6} + \frac{142675}{512} \et^{8} + \frac{825}{256} \et^{10} + \left(\frac{60001}{672} + \frac{2490193}{2688} \et^{2} + \frac{4467613}{2688} \et^{4}\right. \right. \nn\\
& \left.\left. + \frac{1610925}{7168} \et^{6}  - \frac{1565}{32} \et^{8} - \frac{165}{128} \et^{10} \right)\nu  + \left(- \frac{565}{8} - \frac{6105}{8} \et^{2} - \frac{267395}{192} \et^{4} - \frac{53225}{96} \et^{6} - \frac{7175}{256} \et^{8}\right)\nu^{2}\right] \biggr\} \nn\\
& + \nu\,\mutp  \bigg\{ \frac{406211}{3024} - \frac{3236929}{672} \et^{2} - \frac{1234694567}{145152} \et^{4} + \frac{72952835}{3456} \et^{6} + \frac{547268129}{36864} \et^{8}   + \frac{49623629}{36864} \et^{10} + \frac{767205}{114688} \et^{12} \nn\\
& + \left[\frac{8849}{48} + \frac{9309}{4} \et^{2} - \frac{39113257}{1792} \et^{4} - \frac{1031200735}{16128} \et^{6} - \frac{3884615761}{129024} \et^{8} - \frac{1280841431}{516096} \et^{10} - \frac{765435}{57344} \et^{12} \right]\nu \nn\\
& + \left[- \frac{275}{6} + \frac{72181}{36} \et^{2} + \frac{3791191}{192} \et^{4} + \frac{1630667}{48} \et^{6} + \frac{130374383}{9216} \et^{8} + \frac{7114411}{6144} \et^{10} + \frac{3225}{512} \et^{12}\right]\nu^{2} \nn\\
& + \sqrt{1-\et^{2}}\left[- \frac{1880083}{9072} + \frac{11431109}{3024} \et^{2} + \frac{348230363}{18144} \et^{4} + \frac{346072001}{24192} \et^{6} + \frac{298945979}{193536} \et^{8} + \frac{734819}{32256} \et^{10} \right. \nn \\
& \left. + \left(- \frac{118607}{1008} - \frac{3771113}{672} \et^{2} - \frac{42868853}{2016} \et^{4} - \frac{80317183}{5376} \et^{6} - \frac{5098195}{3072} \et^{8} - \frac{67163}{2688} \et^{10} \right)\nu + \left(\frac{505}{4} + \frac{59539}{24} \et^{2} \right. \right. \nn \\\
& \left. \left. + \frac{356347}{48} \et^{4} + \frac{308645}{64} \et^{6} + \frac{460877}{768} \et^{8} + \frac{1687}{192} \et^{10} \right)\nu^{2}\right] \bigg\} + \Big(\sigmatp+\delta\sigmatm\Big) \bigg\{- \frac{173}{756} - \frac{4339}{252} \et^{2} - \frac{142337}{672} \et^{4} \nn\\
&  - \frac{71983}{144} \et^{6} - \frac{134205}{512} \et^{8} - \frac{34969}{1536} \et^{10} + \frac{635}{12288} \et^{12}  + \sqrt{1-\et^{2}}\left(\frac{476}{3} + \frac{12733}{12} \et^{2} + \frac{14637}{8} \et^{4} + \frac{56525}{96} \et^{6} \right. \nn\\
& \left.  + \frac{4165}{96} \et^{8} \right)\nu + \left(- \frac{3262}{27} + \frac{42403}{54} \et^{2} + \frac{554629}{108} \et^{4}  + \frac{1124389}{216} \et^{6} + \frac{21227659}{13824} \et^{8} + \frac{1722973}{13824} \et^{10} + \frac{1365}{1024} \et^{12} \right)\nu\bigg\} \nn\\
& + \nu\sigmatp \biggl\{ - \frac{41389}{378} - \frac{215657}{126} \et^{2}  + \frac{2679535}{112} \et^{4} + \frac{13418269}{288} \et^{6} + \frac{22365341}{1344} \et^{8} + \frac{25715741}{21504} \et^{10} + \frac{91705}{21504} \et^{12} \nn\\
& + \left[\frac{1562}{3}  - \frac{23440}{9} \et^{2} - \frac{1116125}{36} \et^{4} - \frac{3269639}{72} \et^{6} - \frac{9152077}{576} \et^{8} - \frac{2872531}{2304} \et^{10} - \frac{1465}{192} \et^{12} \right]\nu \nn \\
& + \sqrt{1-\et^{2}}\left[\frac{5039}{42} + \frac{558181}{84} \et^{2} + \frac{3656237}{168} \et^{4} + \frac{9253945}{672} \et^{6} + \frac{3969475}{2688} \et^{8} + \frac{1505}{48} \et^{10} \right. \nn \\
& \left. + \left(- \frac{2186}{3} - \frac{24842}{3} \et^{2} - \frac{115189}{6} \et^{4} - \frac{66755}{6} \et^{6} - \frac{113795}{96} \et^{8} - \frac{665}{24} \et^{10} \right)\nu\right]\bigg\}\nn \\
& + \nu\,\muthp \bigg[ - \frac{70}{3} + \frac{3545}{6} \et^{2} + \frac{64465}{32} \et^{4} + \frac{3610}{3} \et^{6} + \frac{455735}{3072} \et^{8} + \frac{175}{128} \et^{10} \nn \\ 
& + \sqrt{1-\et^{2}}\left(50 + \frac{2275}{8} \et^{2} + \frac{5475}{16} \et^{4} + \frac{3775}{64} \et^{6} + \frac{175}{64} \et^{8} \right)\bigg] \,,\\
\mathcal{G}_{7.5} =& \Bigl(\mutp+\delta\,\mutm\Bigr)\bigg[ - \frac{88}{21} + \frac{12429}{32}\et^{2} + \frac{4643039}{896}\et^{4} + \frac{4622424643}{387072}\et^{6} + \frac{6897529229}{1032192}\et^{8} + \frac{995265148289}{1238630400}\et^{10}\nn\\
& + \frac{132364101589}{14863564800}\et^{12} - \frac{76136256817}{7768689868800}\et^{14} -\nu\left( \frac{351}{16} + \frac{6243}{14}\et^{2} + \frac{54374643}{7168}\et^{4} + \frac{722601559}{43008}\et^{6} \right. \nn\\
& \left. + \frac{11405618027}{1179648}\et^{8} + \frac{836568683887}{619315200}\et^{10} + \frac{2362533500837}{33973862400}\et^{12} + \frac{275335017377557}{7768689868800}\et^{14} \right) \bigg] \nn\\
&  +\nu\,\mutp \bigg[  \frac{1159}{168} + \frac{2800753}{672}\et^{2} + \frac{1453557851}{43008}\et^{4} + \frac{8416316141}{193536}\et^{6} + \frac{312522370255}{24772608}\et^{8} - \frac{152760909857}{176947200}\et^{10}\nn\\
& - \frac{160179282325787}{158544691200}\et^{12} - \frac{2589800088115493}{3884344934400}\et^{14} -\nu\left( \frac{2029}{12} + \frac{387199}{42}\et^{2} + \frac{19592705}{512}\et^{4} \right. \nn\\
&\left. + \frac{7549814647}{193536}\et^{6} + \frac{4347049847}{442368}\et^{8} - \frac{95118716269}{103219200}\et^{10} - \frac{34594884376853}{39636172800}\et^{12} \right. \nn\\
& \left. - \frac{6752938011146941}{11653034803200}\et^{14} \right) \bigg] - \Bigl(\sigmatp+\delta\,\sigmatm\Bigr)\bigg[  \frac{2}{9} + \frac{58}{3}\et^{2} + \frac{27385}{144}\et^{4} + \frac{286957}{648}\et^{6} + \frac{10492679}{36864}\et^{8} \nn\\
& + \frac{368318011}{8294400}\et^{10} + \frac{23394227}{31850496}\et^{12} + \frac{6831403}{812851200}\et^{14} \bigg] +\nu\,\sigmatp \bigg[ \frac{904}{9} + \frac{19535}{3}\et^{2} + \frac{173332}{9}\et^{4} \nn\\
& + \frac{70050655}{5184}\et^{6} - \frac{1086311}{9216}\et^{8} - \frac{2976554663}{1843200}\et^{10} - \frac{16728523907}{19906560}\et^{12} - \frac{57801278407409}{104044953600}\et^{14} \bigg] \,.
\end{align}
\end{subequations}

\subsection{Secular evolution of orbital elements at leading order}\label{subsec:SecExpr}

In this section, we write the evolution equations of the secular part of the orbital elements. We only display the leading order for the sake of space, since they are as long as the radiated fluxes~\eqref{eq:FGtot} but we recall that the ancillary file contains the full expressions at relative 2.5PN
\begin{align}
\langle \dot{n} \rangle =& \frac{96 \, x^{11/2}\nu c^6}{5G^2\tmass^2(1-\et^2)^{7/2}}\left(1+\frac{73}{24} \et^2+\frac{37}{96}\et^4\right)\nn\\
& + \frac{576\, x^{21/2} c^6}{5G^2\tmass^2(1-\et^2)^{17/2}} \Biggl\{\Bigl(\mutp+ \delta\,\mutm \Bigr)\left[ 1 + \frac{211}{8}\et^2+\frac{3369}{32}\et^4+\frac{6275}{64}\et^6+\frac{10355}{512}\et^8 + \frac{225}{512}\et^{10}\right] \nn \\
&  \qquad \qquad \qquad \qquad \qquad + \nu \, \mutp \left[-\frac{11}{2} + \frac{1481}{16}\et^2+\frac{8939}{32}\et^4+\frac{15793}{96}\et^6+\frac{13883}{1024}\et^8 \right. \\
& \qquad \qquad \qquad \qquad \qquad \qquad \left. + \sqrt{1-\et^2}\left( -5 + \frac{1237}{24}\et^2+\frac{3869}{64}\et^4+\frac{1813}{192}\et^6 -\frac{29}{128}\et^8\right) \right]\Biggr\}+\calO\left( \frac{1}{c^5}, \frac{\etidal}{c^5}\right)\,,\nn\\
\langle \dot{a}_r \rangle =& - \frac{64  x^{3}\nu c}{5(1-\et^2)^{7/2}}\left(1+\frac{73}{24} \et^2+\frac{37}{96}\et^4\right)\nn\\
& - \frac{384\, x^{8} c}{5(1-\et^2)^{17/2}} \Biggl\{\Bigl(\mutp+ \delta\,\mutm \Bigr)\left[ 1 + \frac{211}{8}\et^2+\frac{3369}{32}\et^4+\frac{6275}{64}\et^6+\frac{10355}{512}\et^8 + \frac{225}{512}\et^{10}\right] \nn \\
&  \qquad \qquad \qquad \qquad \qquad + \nu \, \mutp \left[15 + 142\et^2 +\frac{64877}{192}\et^4+\frac{16915}{96}\et^6+\frac{57259}{3072}\et^8 \right. \\
& \qquad \qquad \qquad \qquad \qquad \qquad \left. + \sqrt{1-\et^2}\left( 3 + \frac{369}{8}\et^2+\frac{429}{8}\et^4+\frac{1649}{128}\et^6 + \frac{37}{64}\et^8\right) \right]\Biggr\}+\calO\left( \frac{1}{c^5}, \frac{\etidal}{c^5}\right)\,,\nn\\
\langle \dot{k} \rangle =& \frac{3c^3\nu x^9 \mutp}{2G\tmass(1-\et^2)^{15/2}}\left( 2880 +4104\et^2+2595\et^4+\frac{267}{2}\et^6\right)+\calO\left( \frac{1}{c^5}, \frac{\etidal}{c^5}\right)\,.
\end{align}
The quantities $x$ and $\et$ in the right-hand-side of these expressions have to be understood as their secular part.

\bibliography{RefList_EccentricTides}

\end{document}